\documentclass[aps,pra,twocolumn,showpacs]{revtex4-1}
\usepackage{amsmath,amssymb,mathrsfs}
\usepackage{graphicx}
\usepackage{amsmath}
\usepackage{epsfig}
\usepackage[usenames]{color}
\usepackage{amsmath}
\usepackage{amssymb}
\usepackage{times}
\usepackage{graphicx}
\usepackage{cancel}

\newcommand{\beq}{\begin{equation}}
\newcommand{\eeq}{\end{equation}}

\definecolor{myblue}{rgb}{0.2,0.2,0.8}
\definecolor{myred}{rgb}{1,0.,0.3}

\begin{document} 

\title{
String tension and robustness of confinement properties of in the
Schwinger-Thirring model}

\author{Joao C. Pinto Barros}
\affiliation{Institute for Theoretical Physics, Bern University, Sidlerstrasse 5, CH-3012 Bern, Switzerland}
\affiliation{SISSA and INFN, Sezione di Trieste, Via Bonomea 265, I-34136 Trieste, Italy}

\author{Marcello Dalmonte}
\affiliation{Abdus Salam International Center for Theoretical Physics, Strada Costiera 11, Trieste, Italy}
\affiliation{SISSA, Via Bonomea 265, I-34136 Trieste, Italy}

\author{Andrea Trombettoni}
\affiliation{CNR-IOM DEMOCRITOS Simulation Center, Via Bonomea 265, I-34136 Trieste, Italy}
\affiliation{SISSA and INFN, Sezione di Trieste, Via Bonomea 265, I-34136 Trieste, Italy}

\begin{abstract}
Confinement properties of the $1+1$ Schwinger model can be 
studied by computing the string tension between two charges. It is  
finite (vanishing) if the fermions are massive (massless) 
corresponding to the occurrence of confinement (screening). 
Motivated by the possibility of experimentally simulate 
the Schwinger model, we investigate here the robustness of its screened and confined phases. Firstly, we analyze
the effect of nearest-neighbour 
density-density interaction terms, 
which -- in the absence of the gauge fields -- give rise to the 
Thirring model. The resulting Schwinger-Thirring model is studied, 
also in presence of a topological $\theta$ term, showing that the 
massless (massive) model remains screened (confined) and 
that there is deconfinement only for $\theta=\pm\pi$ in the massive case. 
Estimates of the 
parameters of the Schwinger-Thirring model are provided with a discussion of a possible experimental setup for its realization with ultracold atoms. 
The possibility that the gauge fields  
live in higher dimensions while the fermions remain in $1+1$ is also considered. 
One may refer to this model as 
the {\it Pseudo}-Schwinger-Thirring model. It is shown that the screening 
of external charges occurs for $2+1$ and $3+1$ gauge fields, 
exactly as it occurs in $1+1$ dimensions, with a 
radical change of the long distance interaction induced by the gauge fields. 
The massive (massless) model continues to exhibit confinement (screening), 
signalling that it is the dimensionality of the matter fields, 
and not of the gauge fields to determine confinement properties. 
A computation for the string tension is presented in perturbation theory. 
Our conclusion is that $1+1$ models    
exhibiting confinement or screening -- massless or massive, in presence of 
a topological term or not -- retain their main properties when the Thirring interaction 
is added or the gauge fields live in higher dimension.
\end{abstract}

%
%

\maketitle

\section{Introduction}
\label{sec:Intro}

The study of confinement properties in gauge theories is 
a long-lasting subject of research, with applications in a variety 
of physical systems ranging from Quantum Chromodynamics (QCD) \cite{Wilczek2000} 
to effective gauge theories emerging 
in strongly correlated systems \cite{Lee2006}. 
The origin of the concept of confinement is associated to the 
fact that no quarks appear as asymptotic particle states. 
Instead, they seem to be confined inside hadrons and mesons. 
At date, there is not yet final and detailed  
description of confinement in QCD. Such difficulties 
are due to the complexity of QCD, a strongly 
coupled non-Abelian gauge theory with symmetry group $SU(3)$. 
Due to this reason, historically, an important role 
in the understanding of confinement was played by solvable 
theories in $1+1$ dimensions - 
an archetypical example being the Schwinger model \cite{Schwinger1962}. This 
is a well studied field theory \cite{zinnjustin}, where 
relativistic fermions are coupled to a $U(1)$ gauge field. It  
exhibits confinement of the fermions and from this point 
of view can be seen as a toy model for QCD \cite{Frishman2010}.

The Schwinger model and its multi-flavour generalization can be mapped 
by bosonization \cite{GNT} onto massive sine-Gordon models,  
having a mass proportional to the fermion charge, but frequency $\beta$ 
fixed to $\sqrt{4\pi}$ \cite{COLEMAN1975,COLEMAN1976} 
(see more references in \cite{NANDORI2008}). 
In the regime of vanishing charge, and upon addition of an interaction term between the fermions, one obtains a (massless) sine-Gordon model 
with variable frequency 
\cite{Coleman1975bis}. This model is the Thirring model \cite{THIRRING1958}. 
In the massless limit its correlation functions are known 
\cite{Hagen1967,Klaiber1969}. In the massive case it is solvable 
by Bethe ansatz \cite{Korepin}.

Both Schwinger and Thirring models separately 
have been heavily investigated for their relevance 
as a toy model for phenomena occurring in higher dimensions 
and for the appealing possibility to study solvable/integrable interacting 
models which may exhibit confinement. In comparison, the model in which $1+1$-d fermions 
are simultaneously charged and interact  
via local interactions -- a model to which one may refer as 
the Schwinger-Thirring model -- has been 
relatively less investigated 
\cite{COLEMAN1975,Frohlich1975,Frohlich1976,dutra1997,BELVEDERE2002,Dalmazi2007,Nandori2010}. 
In spite of the fact that, to the best of our knowledge, 
the Schwinger-Thirring model is not integrable, rigorous results 
on the mapping on the massive sine-Gordon model have been established 
\cite{Frohlich1976}. 

An important way to characterize the confinement properties of models 
such as the Schwinger and the Thirring one is provided by the 
study of the string tension via the determination of the 
energy of the configuration of two probe charges \cite{zinnjustin}. 
More precisely, one has to study how the energy $E$ is modified 
when the two static charges are taken further and further apart. 
If the energy is divergent the theory is said to be {\it confining}. 
The prefactor of the linear term of the energy with respect to 
the distance $L$, with $E \propto L$ for large $L$, 
is called string tension. Due to the phenomenon known as string breaking, 
some models which have no free charged states (under the gauge group) 
may exhibit finite energy for arbitrarily separated charges. 
The final result, is therefore, a {\it screening} of the charges. 
For this reason sometimes is preferred to define confinement 
as the absence of charged asymptotic states. In this paper this 
position will not be adopted and the phases 
will be referred to be either in a confined or screened phase. 
For a general introduction on the confinement problem see, {\it e.g.}, 
\cite{Greensite2011}.

In the $1+1$-d Schwinger model, one can compute 
the string tension between two charges. 
It is found to be finite if the fermions are massive and vanishing if they are massless 
\cite{zinnjustin,Frishman2010}. It is therefore concluded that for massive fermions  one has confinement, and for massless fermions there is the occurrence of screening \cite{Frishman2010}. 

This paper focuses on the confinement properties 
of the Schwinger-Thirring model 
in order to investigate their robustness and 
the role played by the one-dimensionality of the fermions. 
By studying the string tension, the effect 
of the Thirring interactions is addressed first. Secondly, the system where the fermions live in one spatial dimension and 
the gauge fields are defined in $D=2,3$ dimensions is explored. 

This work is directly motivated by the following two reasons. From one side it is intended to determine whether a $1+1$ model   
exhibiting confinement or screening -- massless or massive, in presence of 
a topological term or not -- maintains its properties when interaction 
is added and especially when the gauge fields are allowed to 
live in higher dimension. The latter question is especially relevant 
since the confinement property of the Schwinger model 
could be intuitively explained by the fact that, at the classical level, 
the energy between two-point particles grows linearly with the distance 
for a gauge field in $1+1$ dimensions. When na\"{\i}vely applied, 
this argument would lead to conclude that there should be 
deconfinement when the gauge fields are 
living in three dimensions. 
Therefore, to test the robustness of the confined phase, 
the case where gauge fields are living in higher dimensions ($2+1$ and $3+1$) 
while the fermions remain in $1+1$ is studied. This is the equivalent 
for the Schwinger model of the Pseudo-QED, in which 
the fermions are confined in a $2D$ plane 
while interacting with the electromagnetic field 
living in the $3D$ space 
\cite{Marino93,marino2015}. The Schwinger-Thirring model 
in which the gauge fields live in higher dimension is referred as the 
{\it Pseudo}-Schwinger-Thirring model.

The second motivation for our study comes from recent 
theoretical \cite{tagliacozzo13,zohar2013quantum,zohar15,banerjee12,banerjee2013atomic,hauke2013quantum,stannigel14,bazavov2015gauge,notarnicola15,wiese14,Dalmonte2016LGTSimulations,kasper16} and experimental ~\cite{martinez2016} 
progress on the 
emulation of gauge theories with ultracold atoms and trapped ions. 
In general, quantum simulators \cite{feynman1982simulating,Cirac2012} 
are built up to emulate quantum mechanical systems by 
properly shaping the system dynamics using external fields. Due to the general complexity of the many problem of quantum mechanics, a quantum simulator could provide answers
to long-standing problems in physics. 
The realization of quantum simulations of interacting 
theories could help the understanding of target models, 
the validation of analytical
or numerical techniques, and the exploration of physical phenomena
which are not currently reachable by other approaches. 
Ultracold atoms and trapped ion 
setups provide a variety of reliable different tools 
to perform simulations of different many-body phenomena 
\cite{bloch12,blatt2012quantum}. 

One of the most challenging goals certainly concerns
the implementation of gauge theories  
where non-perturbative 
phenomena, such as confinement \cite{wilson1974confinement}, 
occur. This motivates the study of simpler models which may exhibit
such phenomena. As mentioned earlier, 
the Schwinger model is probably the simplest, non-trivial gauge theory involving fermions one can think of, and, more relevantly, it also exhibits confinement. 
It was also the target of the first quantum simulation of a lattice 
gauge theory addressing the real-time dynamics on a few-qubit 
trapped-ion quantum computer~\cite{martinez2016}. 

Departing from the original model Hamiltonian, two variations are 
experimentally accessible (if not unavoidable) 
according to recent implementation schemes: the 
addition of tunable Thirring interactions, and the engineering 
of a gauge field living in $D=2,3$. As mentioned above, 
these are exactly the two scenarios focused here.
A further interesting ingredient, the topological $\theta$-term, can be added as well. 
In the presence of this extra term, 
deconfinement is possible for $\theta=\pm\pi$, while
the system retains its confining character for any other angle in
between~\cite{COLEMAN1976,Controzzi2003}.  

As previously reminded, it is well known that the massless 
Schwinger model is in the screened phase, while the massive one exhibits confinement \cite{Frishman2010}. 
Here it is shown that for massless fermions the screening phase survives 
when the four-point local interaction term is turned on. 
The confined phase remains present, as well, 
when a finite mass is turned on in the presence 
of interactions, as in the Schwinger model. In both cases 
(massless and massive) 
the string tension does not depend on the Thirring 
interaction coefficient $g$. 
Consistency of the theory requires, for both cases, that $g>-\pi$. 
When the gauge fields are allowed to live in higher dimensions 
($2+1$ or $3+1$), giving rise to the Pseudo-Schwinger-Thirring 
model, the massless (massive) 
model will be shown to remain screened (confined). 
In particular, it is shown that in leading order on the mass, the string tension for gauge 
fields in $1+1$ and $2+1$ dimensions 
is the same. The same scenario occurs in presence of a topological 
$\theta$-term. 
Our results shows that both Thirring interactions and 
gauge fields living in higher dimensions do not alter the confinement 
properties of the Schwinger model with the $\theta$ term, and for 
$\theta=\pi$ the massive (massless) model remains deconfined (screened).

The paper is organized as follows. In Sec.~\ref{sec:implementation}, 
the continuum and lattice formulations of the Schwinger model are introduced, 
briefly discussing the quantum link formulation. The generation 
of a Thirring-like interaction based on nearest-neighbour 
density-density interactions between non-relativistic tight binding 
is presented. In Sec.~\ref{sub_definitions} the definition of the string tension is reminded and the criterion that will be used to determine confinement properties is stated. 
In Sec.~\ref{sec:robustness}, it is shown that 
the confinement properties of the Schwinger model 
are retained under the presence of a Thirring-like interaction. 
In Sec.~\ref{sec:magnitude} an estimate of the parameters of 
the simulated quantum system is provided. This is done using, as reference, the proposal Ref.~\onlinecite{banerjee12}, where a Thirring-like interaction appears naturally as a byproduct of the proposal. The difficulties arising in 
such implementation are discussed and the achievable range of parameters as 
a function of the amplitude of the external superlattice potential is explored. 
Finally, in Sec.~\ref{sec:gauge_in_D}, the Pseudo-Schwinger-Thirring model featuring the gauge fields in higher dimensions is addressed analysing both 
the massless and massive theories. The conclusions are drawn in Sec.~\ref{conclusions}, while additional, more technical 
material is presented in the Appendices.

\section{Implementation and deviations from the Schwinger model}
\label{sec:implementation}

The Hamiltonian for the Schwinger model in the axial gauge $A_{0}=0$ is given by:

\begin{equation}
H=\int dx \left[ \bar{\psi}\left(-i\cancel{\partial}+e\cancel{A}+m\right)\psi+\frac{1}{2}E^2 \right].
\end{equation}

In lattice field theories it is possible to provide a regularization of their continuum counterpart 
by introducing a lattice spacing $a_s$. As a consequence, different lattice formulations may correspond 
to the same model in the continuum limit $a_s\rightarrow 0$. An example is given by 
the Kogut-Susskind (KS) Hamiltonian \cite{kogut1975hamiltonian} which, in the continuum limit for the 
$U\left(1\right)$  gauge theories, gives the QED Hamiltonian in the axial gauge. For the case of 
one spatial dimension, the KS Hamiltonian is given by

$$
H_{KS}=-\frac{i}{2a_s}\underset{n}{\sum}\left(c_{n}^{\dagger}U_{n}c_{n+1}-\mathrm{h.c.}\right)
$$

\begin{equation}
+m\underset{n}{\sum}\left(-1\right)^{n}c_{n}^{\dagger}c_{n}+\frac{a_s}{2}\underset{n}{\sum}E_{n}^{2},
\end{equation}
where $c_n$ annihilates a fermion in the lattice site $n$. 
In the KS model the spinor degrees of freedom are encoded in the spatial
coordinates (staggered fermions)~\cite{Montvay1994} and the continuum limit is taken by sending the lattice spacing to zero $a_s\rightarrow0$. More precisely the continuum variables
are identified through $c_{n}/\sqrt{a_s}\rightarrow\psi_{\mathrm{up}}\left(x\right)$
for $n$ even and $c_{n}/\sqrt{a_s}\rightarrow\psi_{\mathrm{down}}\left(x\right)$
for $n$ odd, to form the two dimensional spinors. $E\left(x\right)$
denotes the electric field. Furthermore the representation
of the gamma matrices is fixed to be $\gamma^{0}=\sigma_{z}$ and $\gamma^{1}=i\sigma_{y}$
where the $\sigma$'s are the Pauli matrices. The gauge operators
obey the commutation relations $\left[U_{m},E_{n}\right]=e\delta_{mn}$
and $\left[U_{m}^{\dagger},L_{n}\right]=-e\delta_{mn}$, with the remaining
commutation relations vanishing. The parameter $e$ is the gauge
coupling. 
The lattice model, besides regularizing the continuum theory, 
provides also a bridge towards the experimental implementation. 

One of the first and well documented difficulties in implementing
such model consists in reproducing the infinite dimensional space spanned
by the gauge fields present on the links (the electric field in each link varies from $-\infty$ to $+\infty$). An alternative approach
consists on truncating the Hilbert space making
it finite and more suitable for quantum simulation. Such models are
known as quantum link models \cite{horn1981finite,orland1990lattice,chandrasekharan1997quantum}. 
In a quantum link model, the link algebra described above is replaced by
the algebra of angular momentum $\left[L_{i,m},L_{j,n}\right]=i\delta_{nm}\varepsilon_{ijk}L_{k,n}$.
Each representation of this algebra, identified by the spin magnitude
$S$, provides a finite Hilbert space. In the limit
of $S\rightarrow+\infty$ the KS Hamiltonian should be recovered.
The commutation
relations between the $U_{m}$'s and $E_{m}$'s are exactly realized 
by identifying $E_{n}\rightarrow eL_{n}$ and $iU_{n}\rightarrow L_{+,n}/S\left(S+1\right)$ 
(where $ L_{+,n}=L_{x,n}+iL_{y,n}$).
However there are now extra non-zero commutation relations: $\left[U_{m},U_{n}^{\dagger}\right]=2\delta_{mn}E_{m}/\left[eS\left(S+1\right)\right]$, reflecting the request that $H_{KS}$ is recovered when $S\rightarrow+\infty$. 
This could constitute a possible drawback on the experimental implementation of
these models. This subject was extensively studied in \cite{kasper16} where
it was shown how the finite dimension of the corresponding Hilbert space may deviate 
from the infinite dimensional case. Qualitatively, 
the finiteness of the quantum links plays a small role and, as expected,
the results converge to the QED result as the value of the total
spin is increased.

Another problem that may emerge is the possibility of nearest-neighbour density-density interaction. In fact, as in \cite{banerjee12}, such term
may be present as a byproduct of the implementation scheme. 
In more general cases, terms of these type may appear quite naturally 
in implementation schemes where gauge invariance is imposed via energetic constraints, due to the fact that such terms are indeed gauge invariant.

From a different perspective, the Schwinger-Thirring model can also host interesting physics which may motivate direct implementations of it in its own. 
In the scheme of Ref.~\cite{banerjee12}, the four-Fermi interaction 
is repulsive and it is not possible to reverse the sign of interaction. The reason 
of such contribution directly comes from the Fermi
statistics. For this reason a possible path towards an implementation where such interactions are tunable may include, for example, an extra species of
bosons with correlated hopping with the fermions through all the lattice,
producing an analogous term with opposite sign. This term could give
extra control on the sign and strength of this interaction.

Building on this intuition, the consequences of the
presence of such term in the gauge theory are investigated. This is done in a general setting where it is admitted a $\theta$-term on the Lagrangian. The lattice Hamiltonian is then modified to be

$$
H_{\mathrm{lattice}}=-\frac{i}{2a_s}\underset{n}{\sum}\left(c_{n}^{\dagger}e^{-ia_seA_{n}^{1}}c_{n+1}-\mathrm{h.c.}\right)
$$
\begin{equation}
+m\underset{n}{\sum}\left(-1\right)^{n}n_{n}
+\frac{e^{2}a_s}{2}\underset{n}{\sum}\left(L_{n}-l_{0}\right)^{2}+\frac{2g}{a_s}\underset{n}{\sum}n_{n}n_{n+1}\label{eq:Hlattice}
\end{equation}
where $n_n=c_{n}^{\dagger}c_{n}$. 
The presence of $l_{0}$ corresponds to a background electric field
which in the continuum limit gives rise to the $\theta$ term with
$l_{0}=\theta/2\pi$. The other extra term is a nearest-neighbour
density-density interaction. This term, which is of the form $\lambda {\sum}n_{n}n_{n+1}$, scales with $a_{s}$ 
to get a finite contribution in the continuum limit, as it can be seen by taking into consideration 
that $\underset{n}{\sum}\rightarrow\frac{1}{2a_s}\int dx$
and the four operators $c_{x}$ bring a factor of $a_s^{2}$ forcing
a pre-factor$\sim a_s^{-1}$. In this way the possible terms of the
form $\psi\partial\psi$ are of higher order and the remaining part
in terms of spinor components is 
$\sim a_{s}\lambda\int dx \psi_{\mathrm{up}}^{\dagger}\left(x\right)\psi_{\mathrm{up}}\left(x\right)\psi_{\mathrm{down}}^{\dagger}\left(x\right)\psi_{\mathrm{down}}\left(x\right)$.
When comparing it to the Hamiltonian of the Thirring model $\frac{g}{2}\int dx \left(\bar{\psi}\gamma_{\mu}\psi\right)\left(\bar{\psi}\gamma^{\mu}\psi\right)$,
which results in a term $2g\int dx \psi_{\mathrm{up}}^{\dagger}\left(x\right)\psi_{\mathrm{up}}\psi_{\mathrm{down}}^{\dagger}\left(x\right)\psi_{\mathrm{down}}\left(x\right)$
one concludes that the correspondent lattice parameter is given by $2g/a_s$.

The inclusion of the $l_0$ term, and equivalently of the $\theta$ term, corresponds to a shift of the background 
electric field of the vacuum. This is a consequence of the presence of a linear term $\propto L_n$. 
In the scheme \cite{banerjee12}, that will be discussed in Section \ref{sec:magnitude} for the estimate of 
the achievable range of parameters, this corresponds to create an extra imbalance between 
the chemical potentials of the boson species. 

\subsection{Confinement and screening in $1+1$ gauge theories}
\label{sub_definitions}

Here the existence of confinement is characterized by computing the string tension $\sigma$ between two external charges added to the system. 
Since such characterization of confinement properties in Schwinger-Thirring 
models is the main subject of the present paper, a pause 
is made here to relate confinement, deconfinement and screening to 
the string tension $\sigma$. 
This quantity is defined as the constant
of proportionality between the energy $T$ of two
added external charges and their distance $L$: 
\begin{equation}
T=\sigma L,
\end{equation} 
when, for large $L$, $\sigma$ is positive one has confinement, 
while $\sigma<0$ one has a deconfined phase. In the case in which $\sigma=0$, 
one then studies the limit for large $L$ of the energy $T$: if it is infinite 
and positive/negative then one has respectively 
confinement/deconfinement. When $\sigma$ tends to 
$0$ and $|T|$ is not diverging, then the ratio 
$T_{with}/T_{without}$ can be considered, where $T_{with}$ ($T_{without}$) is the energy with (without) the fermionic fields. If this ratio is vanishing, one has a screened phase, while in the other cases it is not possible to conclude about confinement, deconfinement or screening just looking at the energy $T$. Instead one should look at the behaviour (and the poles) of correlation functions 
in order to determine the presence/absence of charged asymptotic states \cite{zinnjustin}.

This way of characterizing confinement properties covers the cases to be considered here.

\section{Robustness under Thirring interactions}\label{sec:robustness}

The problem of confinement in the Schwinger and Thirring models in $1+1$ dimensions 
can be addressed through bosonization \cite{GNT}. 
The general procedure adopted here follows closely \cite{dutra1997} and it will be be the basis 
for the results presented in Section \ref{sec:gauge_in_D} where the gauge fields are living 
in higher dimensions. 

The continuum Lagrangian in Euclidean time for the Schwinger-Thirring model reads:
\begin{equation}
{\cal L}=-\bar{\psi}\left(\cancel{\partial}+ie\cancel{A}+m\right)\psi+\frac{g}{2}\left(\bar{\psi}\gamma_{\mu}\psi\right)^{2} 
+\frac{1}{4}F_{\mu\nu}^{2}+i\frac{e\theta}{4\pi}\varepsilon_{\mu\nu}F^{\mu\nu}
\label{eq:Lagrangian_STtheta}
\end{equation}
(until differently stated, one uses units where $\hbar=c=1$).

For $g=0$, the model (\ref{eq:Lagrangian_STtheta}) is the Schwinger model 
with the $\theta$-term, which is known to exhibit (partial)
deconfinement only for $\theta=\pm\pi$ \cite{COLEMAN1976,Controzzi2003}. 
In turn, for $e=0$ 
one has the the Thirring model. Such theory makes sense only for $g>-\pi$, 
as discussed below, and it can mapped order by order in the mass $m$ 
to a sine-Gordon model \cite{Coleman1975bis}.  
In the following it is showed that these results remain valid when both
parameters $e$ and $g$ are finite in the Schwinger-Thirring model. In particular the Thirring term does not play
any role on confined/deconfined phases of the model. 

The quartic fermionic interaction in (\ref{eq:Lagrangian_STtheta}) can be re-expressed 
via an Hubbard-Stratonovich transformation. This amounts
to replace in the Lagrangian $\frac{g}{2}\left(\bar{\psi}\gamma_{\mu}\psi\right)^{2}\rightarrow-ieB_{\mu}J_{\mu}+\frac{e^{2}}{2g}B_{\mu}^{2}$, with now the integration being performed over $B_\mu$ as well.
Even though the parameter $e$ is entering the transformation, the integration of the fields $B_{\mu}$ produce a result 
independent of $e$ and such introduction of the matter-gauge coupling is present for convenience only. 
In order to isolate the coupling with the fermion fields the variable transformation $A_{\mu}\rightarrow C_{\mu}=A_{\mu}+B_{\mu}$ is performed. The resulting theory reads:

$$
{\cal L}=-\bar{\psi}\left(\cancel{\partial}+ie\cancel{C}+m\right)\psi+\frac{1}{4}F_{\mu\nu}^{\left(c\right)2}+\frac{1}{4}F_{\mu\nu}^{\left(b\right)2} 
$$

\begin{equation}
-\frac{1}{2}F_{\mu\nu}^{\left(b\right)}F_{\mu\nu}^{\left(c\right)}+i\frac{e\theta}{4\pi}\varepsilon_{\mu\nu}\left(F_{\mu\nu}^{\left(c\right)}-F_{\mu\nu}^{\left(b\right)}\right)+\frac{e^{2}}{2g}B_{\mu}^{2}\label{eq:LagrangianHS2-1}
\end{equation}
where the indices $b,c$ refers respectively to $B$ and $C$ fields. The result is then a standard Schwinger model plus an extra boson field coupled to the
gauge field. It is worth noting that this is a peculiar type of interaction
which is independent of the parameters $e$ and $g$. These parameters
control the interaction between the gauge field and the fermions, and
the mass of the field $B_{\mu}^{2}$. 

The gauge transformations are
encoded in the usual way in the fields $\psi$ and $C_{\mu}$. In
turn the $B_{\mu}$ field does not transform under gauge transformation.
Exploiting gauge freedom, one can pick the Lorentz gauge where $\partial^{\mu}C_{\mu}=0$
and therefore parametrize the field $C$ as as $C_{\mu} \equiv -i\varepsilon_{\mu\nu}\partial^{\nu}\varphi$
which is a general form for divergenceless fields in two dimensions.
The field $B$
has be parametrized with a gradient part too: 
$B_{\mu} \equiv \partial_{\mu}\chi^{\prime}-i\varepsilon_{\mu\nu}\partial_{\nu}\varphi^{\prime}$.
It turns out that the field $\chi^{\prime}$ decouples from the remaining ones and therefore can be left out for this analysis.

Upon performing a change of variables through a chiral transformation
$\psi=e^{ie\varphi\gamma_{5}}\psi^{\prime}$ where $\gamma_{S}=i\gamma_{0}\gamma_{1}$,
the extra term $-\bar{\psi}\left(ie\cancel{\partial}\varphi\gamma_{S}\right)\psi$
can be written as $-\bar{\psi}\left(ie\cancel{\partial}\varphi\gamma_{S}\right)\psi=e\varepsilon_{\mu\nu}\bar{\psi}\gamma_{\nu}\partial_{\mu}\varphi\psi$, so that the coupling term to $C_{\mu}$ is vanishing. 
In turn, the massive term is mapped
to $-m\left(\bar{\psi}\psi\cos2e\mbox{\ensuremath{\varphi}}+i\bar{\psi}\gamma_{5}\psi\sin2e\mbox{\ensuremath{\varphi}}\right)$.
Furthermore, induced by the chiral anomaly, an extra term arises with the form 
$-\frac{e^{2}}{2\pi}\left(\partial_{\mu}\varphi\right)^{2}$.
The Lagrangian on $\psi^{\prime}$ is now decoupled from the rest
of the system and it can be mapped to the bosonic action $\frac{1}{2}\left(\partial_{\mu}\vartheta^{\prime}\right)^{2}-\mu\cos\left(\sqrt{4\pi}\vartheta^{\prime}+2\varphi\right)$
where $\mu=me\exp\left(\gamma\right)/\left(2\pi^{3/2}\right)$. Translating
the $\vartheta^{\prime}$ through $\vartheta^{\prime}=\vartheta-\frac{e}{\sqrt{\pi}}\varphi$
the full Lagrangian is then quadratic on $\varphi$ and $\varphi^{\prime}$
and it can be integrated out. 

The full Lagrangian after this procedure
is given by:

$$
{\cal L}=\frac{1}{2}\left(\partial_{\mu}\vartheta\right)^{2}-\frac{e}{\sqrt{\pi}}\partial_{\mu}\vartheta\partial_{\mu}\varphi-\mu\cos\left(\sqrt{4\pi}\vartheta\right) 
$$

$$
-\frac{1}{2}\left(\partial^{2}\varphi\right)^{2}-\frac{1}{2}\left(\partial^{2}\varphi^{\prime}\right)^{2}
+\partial^{2}\varphi^{\prime}\partial^{2}\varphi
+\frac{e\theta}{2\pi}\left(\partial^{2}\varphi-\partial^{2}\varphi^{\prime}\right)
$$
\begin{equation}
+\frac{e^{2}}{2g}\left(\left(\partial_{\mu}\chi^{\prime}\right)^{2}-\left(\partial_{\mu}\varphi^{\prime}\right)^{2}\right)
\end{equation}

The field $\chi^{\prime}$ only appears in the kinetic term 
and is not coupled to the other fields as announced before. 
The $\theta$-term of $\varphi$ can be re-cast in cosine form by transforming
$\vartheta\rightarrow\vartheta+\theta/\sqrt{4\pi}$. The integration
over $\varphi$ therefore gives 
a term of the form $-\frac{1}{2}\left(\frac{e}{\sqrt{\pi}}\vartheta+\partial^{2}\varphi^{\prime}\right)^{2}$.
Crucially, the coupling between $\varphi$ and $\varphi^{\prime}$
has just a prefactor $1$ in front of it, making the terms $\left(\partial^{2}\varphi^{\prime}\right)^{2}$
cancel out. 
This result is due to the fact that one of the gauge fields is actually
a fictitious field derived from a Thirring interaction which finally controls its mass 
and not the interaction with the other
gauge field. 

The next step consists in integrating out the field $\varphi^{\prime}$.
This field is also related to a $\theta$-term with the opposite sign, 
inducing a new transformation $\vartheta\rightarrow\vartheta-\theta/\sqrt{4\pi}$ (opposite to the one made above). 
However, the $\vartheta$ field acquired a mass, so as expected the dependence
on $\theta$ is not erased but instead is explicit in the term 
$\frac{e^{2}}{2\pi}\left(\vartheta-\theta/\sqrt{4\pi}\right)^{2}$.
Nonetheless this transformation is useful to perform the integration
on the $\varphi^{\prime}$ field, which appears in the form $\frac{e^{2}}{2g}\left(\partial_{\mu}\varphi^{\prime}\right)^{2}-\frac{e}{\sqrt{\pi}}\partial_{\mu}\vartheta\partial_{\mu}\varphi^{\prime}$.
The integration brings the Thirring contribution to the bosonic action
$\frac{g}{2\pi}\left(\partial_{\mu}\vartheta\right)^{2}$. Finally,
the usual mass term is restored by the transformation $\vartheta\rightarrow\vartheta+\theta/\sqrt{4\pi}$. 
The Lagrangian finally reads:

\begin{equation}
{\cal L}=\frac{1}{2}\left(1+\frac{g}{\pi}\right)\left(\partial_{\mu}\vartheta\right)^{2}+\frac{e^{2}}{2\pi}\vartheta^{2}-\mu\cos\left(\sqrt{4\pi}\vartheta+\theta\right)\label{eq:Bosonized_STtheta}.
\end{equation}
The Lagrangian (\ref{eq:Bosonized_STtheta}) shows the contribution 
of both the Schwinger and the Thirring models. No mixing between the
$g$ and $e$ couplings occur. It is worth noting that the above Lagrangian
corresponds exactly to the Lagrangian obtained from (\ref{eq:Lagrangian_STtheta})
if one applies the mapping obtained from bosonization of the free massless
Dirac fermions integrating out the gauge field. More precisely, this
amounts to replace the pure fermionic action by the respective sine-Gordon
model and replace the current by 
$\bar{\psi}\gamma_{\mu}\psi=\frac{1}{\sqrt{\pi}}\varepsilon_{\mu\nu}\partial_{\nu}\vartheta$
both in the coupling to $A_{\mu}$ and in the Thirring term contribution.
After translating the bosonic field $\vartheta$ by $\theta/\sqrt{4\pi}$
and integrating the gauge field, the Lagrangian (\ref{eq:Bosonized_STtheta})
is obtained. The fact that such substitution holds is not obvious 
and it guarantees that the final result is correct.

From the bosonized Lagrangian (\ref{eq:Bosonized_STtheta}) it follows 
that, as in the Thirring model, it is required $g>-\pi$ in order the model itself makes sense. The existence
of a threshold for a minimum of $g$ is expected from the lattice 
theory. For $g\rightarrow-\infty$ the dominant term on the Hamiltonian
is a nearest-neighbour strong repulsion which will induce a phase
separation in the system. For that limit the field theory, or, in other
words, the continuum limit, cannot be taken. 

Regarding the screening and confinement properties of the model, the simplest case corresponds to the
massless theory where $\mu=0$. The propagator for the bosonic
theory is given by

\begin{equation}
\label{eq:propagator}
\Delta_{\vartheta}(p)=\frac{1}{\left(1+\frac{g}{\pi}\right)p^{2}+\frac{e^{2}}{\pi}}.
\end{equation}
From (\ref{eq:propagator}) one can compute, for instance, the two-point function of the scalar
$\bar{\psi}\psi$ and pseudoscalar $\bar{\psi}\gamma_{S}\psi$ fields, 
showing that there are no charged fermions in the spectrum. The derivation of this result can be found in the 
Appendix \ref{sec:2pntfunctions}.

For the massive case a perturbative analysis can be found on Appendix \ref{sec:massive_tension} 
where the string tension is computed. The computation of the string tension for the massless case of 
the Schwinger model is treated in Section \ref{sec:gauge_in_D}  as a particular case of the general 
construction with gauge fields in $D+1$ dimensions. The screening of the external charges becomes 
then explicit with an exponential decay of the energy with the distance between charges. 
As a consequence, the string tension is vanishing.

The results of this Section, and in particular the form of the Lagrangian (\ref{eq:Bosonized_STtheta}), 
clearly show that the confinement and screening phases of the Schwinger model, respectively for the massive and massless cases, are not altered by the Thirring interactions terms. In the next Section an estimate of the lattice parameters of the Schwinger-Thirring model in setups of ultracold atoms is provided, while in Section \ref{sec:gauge_in_D} the robustness of confinement when the gauge fields live in higher dimensions is discussed.

\section{Estimates of lattice parameters for ultracold atom 
setups} \label{sec:magnitude}

A number of proposals have been put forward to simulate the Schwinger model. These include both proposals 
in which the gauge symmetry emerges
as a symmetry of the low energy effective theory \cite{banerjee12,kapit2011optical,notarnicola15} or
as an exact symmetry of the system \cite{zohar2013quantum,kasper16}. An important point to be stressed 
is that fine tuning of parameters should be neither required or crucial since small 
inaccuracies in the experimental implementation could be spoil the validity of the quantum simulation.
In this Section the issue of the deviations from the desired Hamiltonian due to 
the presence of a nearest-neighbour density-density interaction is discussed. It is argued that the presence of 
such four fermion interaction term corresponds to a  Thirring interaction. A scheme for simulating the Thirring 
model was also put forward in \cite{cirac2010cold}. 

\begin{figure}
\begin{centering}
\includegraphics[scale=0.35]{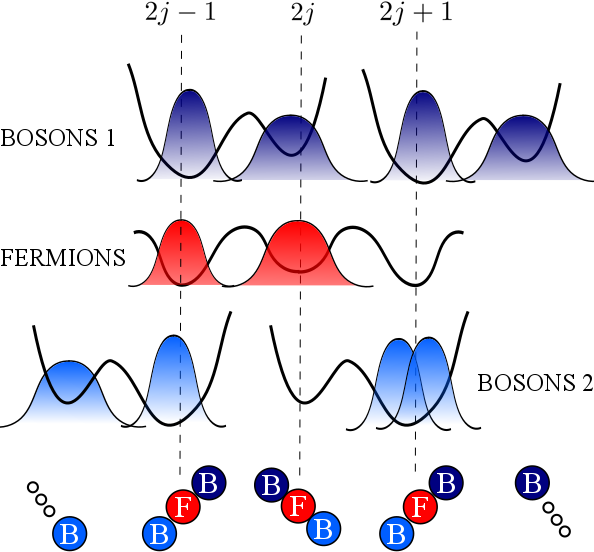}
\caption{\label{fig:scheme}Superlattice configurations for two boson species and the fermionic one. 
Bosons of the species $1$ at an even site $2j$ can only hop to $2j-1$ while a boson of species $2$ has 
only access to the site $2j+1$. The figure presents a an example of a gauge invariant state 
configuration (on these three sites) where $G_{n}\left|\psi\right\rangle =0$}. 
\par\end{centering}
\end{figure}

This Section also provides an estimate of the values
of a different parameters in a possible implementation with optical lattices 
\cite{Lewenstein2012}. 
The main goal here is to provide a quantitative estimate of the energy scales 
in the system, and in particular, of the role of Thirring terms.
To this end, and in order to have a specific example at hand, 
the model described in \cite{banerjee12} was chosen. There, the density-density interaction appears explicitly. This model makes use of one species of fermions and two species of bosons and it builds the quantum links using the Schwinger representation. The fermions are hopping
between all lattice sites while odd links are associated with one
species of bosons, and even links with the other. Each boson is then
only allowed to hop between its designated link. In concrete, 
at any lattice site $n$, one can have bosons of both species.
If $n$ is odd then the species $1$ can only hop to $n+1$ and the species
$2$ to site $n-1$. The opposite happens for $n$ even. This 
is illustrated in Figure \ref{fig:scheme} where both species of bosons and fermions are represented 
around an even site. 

The quantum links are realized through the Schwinger 
representation where 
$$L_{z,n}=\frac{1}{2}\left(b_{n+1}^{\left(\sigma\right)\dagger}b_{n+1}^{\left(\sigma\right)}-b_{n}^{\left(\sigma\right)\dagger}b_{n}^{\left(\sigma\right)}\right),$$
with $\sigma$ indicating the bosonic species ($1$ or $2$) that has to be 
consistent with the parity of the link in question. In this language
the generator of gauge transformations can be written as 
$$G_{n}=n_{n}^{F}+n_{n}^{1}+n_{n}^{2}-2S+\frac{1}{2}\left(\left(-1\right)^{n}-1\right),$$
where $n_{n}^{F}$ indicates the number of fermions and $n_{n}^{\sigma}$
the number of bosons of species $\sigma$ in the site $n$. Therefore if
writing an Hamiltonian of the form $H=H_{0}+U\sum G_{n}^{2}$ where
the energy scale $U$ is much larger than the energy scales of $H_{0}$,
one always obtains, in perturbation theory, a gauge invariant low energy Hamiltonian. 
When considering $H_{0}$ to be the sum
of single particle Hamiltonians of the three considered species, the effective Hamiltonian is of the form (\ref{eq:Hlattice}). 

Adopting the typical notation used for ultracold atoms \cite{Lewenstein2012}, 
one has to evaluate the hopping parameters 
$t_{\alpha}$ of the (non-relativistic) ultracold atom mixture and the 
parameters of the lattice Hamiltonian (\ref{eq:Hlattice}).
The index $\alpha$ is here used to denote 
the fermions or one of the two bosonic species: 
$\alpha\in\left\{ F,1,2\right\}$. To establish the connections 
between the parameters $t_{\alpha}$ and the 
lattice Hamiltonian (\ref{eq:Hlattice}), one has to restore $\hbar$
and $c$ in the Hamiltonian, which corresponds to add $\hbar c$ to
all terms except the mass term which gets a $c^{2}$. The parameters
of the KS Hamiltonian are $a_{s}$ (which is {\it not} the lattice spacing
of the cold atomic optical lattices), the electric charge $e$ and the Thirring
term $g$. The kinetic term is characterized by $t_{B}t_{F}/U$, the
pure bosonic term by $t_{B}^{2}/U$ and the nearest-neighbour density-density
term by $t_{F}^{2}/U$. 
The connection with the parameters of the KS Hamiltonian is given by 
$$4g=-\frac{t_{F}}{t_{B}}$$ and $$e^{2}a_{s}^{2}=\frac{t_{B}}{2t_{F}},$$
from which one conclude that the parameters cannot be varied independently. 

\begin{figure}
\begin{centering}
\includegraphics[scale=0.33]{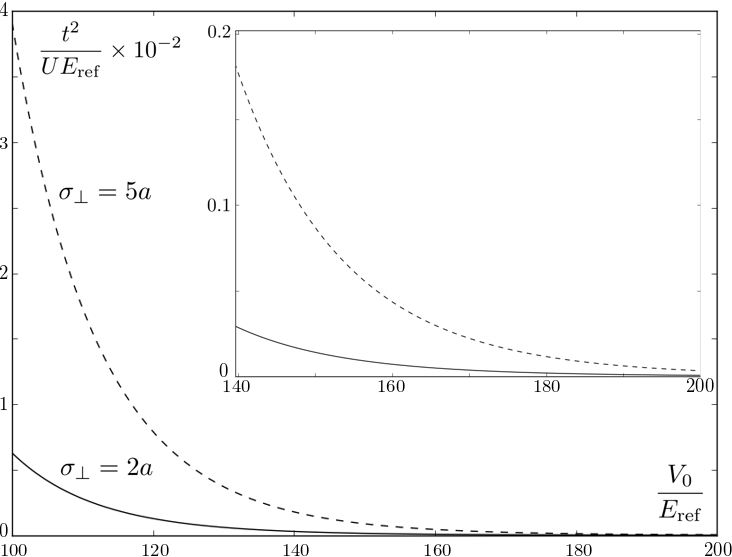}
\caption{\label{fig:plots} Values of the energy scales of the effective Hamiltonian parameters, which collapse in a single curve at this scale,
as the amplitude $V_0$ of the external potential is varied. Two cases, corresponding to $\sigma_\perp=2a$ and $\sigma_\perp=5a$, are presented.
The energies are measured in units of $E_{\mathrm{ref}} \equiv \hbar^{2}/2ma^2$, where $a$ is the lattice spacing. A zoom of the plot is also presented highlighting a  range of values where one enters the desired perturbative regime.}
\par\end{centering}
\end{figure}

A key and challenging aspect of this implementation scheme is that the interactions between the different atomic species shall satisfy the conditions 
$U_{11}=U_{22} \equiv U$ and $2U_{12}=2U_{1F}=2U_{2F}=U$ \cite{banerjee12}, 
where $U_{\alpha\beta}$
is the on-site interaction parameter between species $\alpha$ and
$\beta$ \cite{Lewenstein2012}. The related problem is that the 
proposal requires the bosons sitting in asymmetric minima. This
asymmetry may change the Wannier functions entering the 
evaluation of the $U$ parameters \cite{Lewenstein2012}. This 
may lead to interactions of the form $U_{\alpha\beta}^{+/-}$
where the labels $+/-$ indicate the relative ($+$) and absolute ($-$) minima of the optical lattices involved in the experimental implementation. 
However, the existence of different parameters
$U^{+/-}$ is not crucial. Gauge
invariance at low energies is obtained in perturbation theory for large $U$ so as
long as both $U^{+/-}$ remain larger than the other parameters of
the model the condition $G_{x}\left|\psi\right\rangle =0$ still holds.
Since the goal of this Section is to provide an estimate 
of the parameters of the underlying lattice Hamiltonian, this asymmetry is disregarded, since it can be made
small. Details about this point are given in 
Appendix \ref{sec:parameter_estimate}. Furthermore, 
to satisfy the conditions 
$U_{11}=U_{22} \equiv U$ and $2U_{12}=2U_{1F}=2U_{2F}=U$ 
one still has to match the different interactions between
atomic species. Again, the system can be shown to be 
robust under small enough deviations
from these conditions. Possible deviations lead to 
$UG_{x}^{2}\rightarrow UG_{x}^{2}+U\sum\Delta U_{\alpha\beta}
n_{x}^{\alpha}n_{x}^{\beta}$, 
where $\Delta U_{\alpha\beta}$ is the deviation of the interaction
between species $\alpha$ and $\beta$ from the desired value. The fundamental
requirement to obtain a gauge invariant theory is still valid as
long as $\Delta U_{\alpha\beta}/U\ll1$. 

To estimate the lattice parameter, a mixture 
of $^{52}\mathrm{Cr}$ is used as reference.
This particular choice is related to the fact, 
as explained below, that their scattering lengths approximately have 
the required interaction strength between the two bosonic species 
foreseen by the proposal. 
The scattering length between pairs with total angular
momentum equal to $0$ is $a_{b0} \simeq 30-50$  and for pairs with
total angular momentum equal to $4$ is $a_{b4}\simeq58\pm6$, with 
$a_0$ the Bohr radius \cite{werner2005,stamper2013spinor}. With 
$a_{b0}=30a_{0}$ (different species) and $a_{b4}=60a_{0}$ 
(same species) one has the 
required relations between $U_{11}$, $U_{22}$ and $U_{12}$. For
the fermionic species it is assumed that a tuning of the interaction
is possible in such a way that the remaining conditions can be obeyed.
For the present calculation it is assumed, as well, that $a_{1F}=a_{2F}=30a_{0}$.
Notice that the presence of bosonic-bosonic interactions
does not have the fermionic-fermionic counterpart and that fermionic and bosonic densities are different (see the difference between
the two equations in \ref{eq:sdeviations}). However, as long as $S$
is not very large, it is expected that the effect is not significant.
Here the calculations will be performed for $S=1$ and this assumption
proves to be enough. In order to increase $S$
one should carefully adjust $a_{1F}$ and $a_{2F}$. 
Mixtures involving alkaline-earth-like atoms may prove to be 
even more convenient, due to the possibility of easily engineering species 
dependent lattices there.

The Wannier functions are assumed to be Gaussian with a width 
$\sigma_{\alpha}^{i}$, depending, also, on the direction $i$ and determined 
variationally by minimizing the
Gross-Pitaevskii energy 
(see more details in \cite{Trombettoni05,PintoBarros17}). 
The width of the Gaussians in the directions perpendicular 
to the optical lattices is considered to be $\sigma_{\perp}$,
assumed here to be the same for all the species (fermions and bosons).
The potential felt by the particles is characterized by an amplitude
$V_{0\alpha}$, which controls the height of the barrier, and by  $V_{0\alpha}\Delta_{\alpha}$, the off-set controlling the difference of energy
between minima (with lattice spacing $a$). 

It is assumed $S=1$, $V_{0F}=V_{0B}=V_{0}$ and $\Delta_{F}=\Delta_{B}=0$.
For the lattice spacing it is taken $a\sim 0.5 \mu m$. The process is detailed in Appendix
\ref{sec:parameter_estimate}. In the following, we specifically address the case of a 1D optical lattice, where transverse confinement in the direction perpendicular to the wire is shallow. Similar ideas are applicable in the 3D optical lattice case. Two different choices for $\sigma_\perp$ were considered: $\sigma_\perp=2a$ and $\sigma_\perp=5a$. For smaller values, the variational approach is not expected to be accurate while, for larger values, very large potential depths are required to enter in the necessary perturbative regime. The main result is illustrated in Fig.~\ref{fig:plots}, where the effective parameters $t_{B}t_{F}/U$, $t_{B}t_{F}/U$ and $t_{B}t_{F}/U$ are plotted in a single curve (one for each $\sigma_\perp$). This is because, at this scale, they are indistinguishable. Furthermore, with these parameters, $t_{\alpha}/U$ becomes approximately $10^{-1}$ at $V_0=130E_\mathrm{ref}$ and $V_0=180E_\mathrm{ref}$, respectively. Note that, when compared to the usual recoiled energy, $E_\mathrm{ref}$ is $\pi^2$ times larger. By adjusting some parameters, other choices are of course possible.
Such values were chosen in order to give an illustrative example. 
For these range of parameters, one finds that the relation $2U_{1F}=2U_{2F}=U$ is satisfied very accurately with an error inferior to $10^{-2}\%$. At the same time, the ratio $t_{F}/t_{B}$ remains always very close to $1$.

In conclusion, the range of parameters one may reasonably have access to
corresponds to
$-g \sim 0.25$ and $e^{2}a_{s}^{2}\sim 0.5$ bounded by the condition
$-8/g=e^{2}a_{s}^{2}$. The typical energy of the processes of the
effective cold atomic system is of the order
$t_{\alpha}t_{\beta}/U,\ V\Delta\sim10^{-3}E_{\mathrm{ref}}$. 
Moreover, in this regime the value of the three terms $t_{\alpha}t_{\beta}/U$ become indistinguishable. Finally it is observed that 
the strength of the mass parameter is less constrained and 
it can be changed via the quantity $\Delta$.  This parameter was set to zero during the parameter estimation but, as long as it is not too large, the approximation remains valid.

\section{Screening with gauge fields in $D+1$ dimensions}\label{sec:gauge_in_D}

In this Section the Schwinger-Thirring model in presence of a gauge field living in higher dimension is considered. 
As briefly mentioned in the Introduction \ref{sec:Intro}, 
na\"{\i}vely one would expect to find 
deconfinement is in $D=3$ spatial dimensions, 
one may 
anticipate to find features of normal QED with the particularity of the ``electrons'' 
being restricted to one spatial dimension. Here it is showed that the situation is more subtle.

Models exhibiting dimensional mismatch 
between relativistic fermions and gauge fields have been studied 
and applied in a variety of situations: 
in the context of graphene 
\cite{marino2015,menezes2016,menezes2016influence} and 
transition-metal dichalcogenides \cite{Nascimento2017,marino2018quantum}, 
for topological insulators both for $2+1$ \cite{menezes2017excitonic} 
and $1+1$ fermions \cite{menezes2016conformal}, for $1+1$ fermionic models 
\cite{Miransky2001,menezes2016conformal,barrosarXiv} and 
as a source to generate effective 
short-range \cite{alves2017two} and long-range \cite{barrosarXiv} 
interactions via the dimensional mismatch.  
Besides the direct relevance of these models to experimental systems, such as graphene in a $3D$ 
electro-magnetic field \cite{Marino93}, 
these models are also promising from the points of view of theoretical study and of experimental implementation 
perspectives. The most immediate application is given by the realization of an intermediate step towards 
the simulation of increasingly complicated gauge theories. For this, the necessary correlated hopping of bosons 
and fermions is restricted to the line, while the gauge fields still live in higher dimensions. 
In terms of implementation complexity, these models are expected to be more complicated to implement 
than the Schwinger model but still simpler than QED. The same applies if the gauge fields are non-Abelian.

In the following the basic construction of theories in which the gauge fields live in $D+1$ dimensions is revised. 
The general form of the Lagrangian is given, in Euclidean time, by:
\begin{equation}
{\cal L}={\cal L}^{d+1}_M-ie j^{\mu}_{D+1}A_{\mu}+\frac{1}{4}F_{\mu\nu}^{2}+{\cal L}_{GF}\label{eq:QEDLagrangian}.
\end{equation}
The fermions are considered to live in the lower dimensionality $d+1$, which is made explicit in the matter Lagrangian ${\cal L}^{d+1}_M$. The gauge field lives in the higher dimensionality $D+1$. 
The $D+1$ current is taken to be:

\begin{equation}
\begin{split}
j_{D+1}^{\mu}\left(x^{\alpha}\right)=\left\{ \begin{array}{ll}
j_{d+1}^{\mu}\left(x{}_{0},\ldots,x_{d}\right)\delta\left(x{}_{d+1}\right)\ldots\delta\left(x_{D}\right),\\
\ \ \ \ \ \ \ \ \ \ \ \ \ \ \ \ \ \ \ \ \ \ \ \ \ \ \ \ \ \ \ \ \ \ \ \ \mathrm{if}\ \mu=0,\ldots,d\\
0\ \ \ \ \ \ \ \ \ \ \ \ \ \ \ \ \ \ \ \ \ \ \ \ \ \ \ \ \ \ \ \ \ \ \ \ \ \ \ \ \ \ \ \mathrm{otherwise}
\end{array}\right.\label{eq:fromDtod}
\end{split}
\end{equation}
In (\ref{eq:QEDLagrangian}) the term ${\cal L}_{GF}$ corresponds to the Fadeev-Popov gauge fixing
term which is given by ${\cal L}_{GF}=\frac{1}{2\xi}\left(\partial_{\mu}A_{\mu}\right)^{2}$, 
where different choices of $\xi$ correspond to different gauges. The Feynman gauge, where $\xi=1$ and one has a diagonal propagator $G_{\mu\nu}=\frac{1}{-\partial^{2}}\delta_{\mu\nu}$, is adopted.

The theory (\ref{eq:QEDLagrangian}) can be suitably formulated only in the lower dimension $d$ 
without explicitly invoking higher dimension. The effect of the higher dimensionality 
is encoded in a modified kinetic term for the gauge fields. For the case of $d=2$ and $D=3$ 
it goes under the name of Pseudo-QED \cite{Marino93}. 
Details on the general construction can be found in  \cite{barrosarXiv}. The general form of the Lagrangian is given by:

\begin{equation}
{\cal L}_{d}={\cal L}^{d+1}_M-ie j^{\mu}_{d+1}A_{\mu}+\frac{1}{4}F_{\mu\nu}\hat{M_D}F_{\mu\nu},\label{eq:LongRange_GaugeFields}
\end{equation}
where the operator $\hat{M_D}$ is given in terms of the propagator $\hat{G_D}$ of the gauge fields according to the 
relation $\hat{M_D}=\left(-\partial^{2}\hat{G_D}\right)^{-1}$. 
In the previous expression $\partial^{2}$ is the Laplacian in $d+1$ 
dimensions. The explicit expression for $\hat{M_D}$ (or $\hat{G_D}$) depends 
on the higher dimension $D$ but all the fields are now exclusively 
in $d+1$ dimensions. 
The calculation of the effect of external charges on the system can be done as for the Schwinger case. 
This amounts to introduce an extra
contribution $-iQA_{\mu}j_{\mathrm{ext}}^{\mu}$ where $Q$ is
the absolute value of the two opposite external charges. This external
current can be written in the form $j_{\ \mathrm{ext}}^{\mu} \equiv \varepsilon_{\mu\nu}\partial_{\nu}K$
and it can be eliminated by a chiral transformation. The variable change
corresponds to $\psi=e^{iQK\varphi\gamma_{5}}\psi^{\prime}$ where, again,
one should take into account the chiral anomaly. Notice that the modified kinetic term of the gauge field has no effect on the procedure. 
Once the gauge fields are integrated out, the resulting bosonic theory is given by:

$${\cal L}=\frac{1}{2}\phi\left(-\partial^{2}+\frac{e^{2}}{\pi}M_D^{-1}\right)\phi-\mu\cos\left(\sqrt{4\pi}\phi\right)$$
\begin{equation}
+\frac{eQ}{\sqrt{\pi}}\phi M^{-1}K+\frac{Q^{2}}{2}KM^{-1}K.
\end{equation}

With the field transformation $$\phi^{\prime}=\phi+\frac{eQ}{\sqrt{\pi}}KM^{-1}\frac{1}{-\partial^{2}+\frac{e^{2}}{\pi}M^{-1}},$$ the coupling between the field $K$ and the bosonic field is translated to a sine-Gordon form. 
The resulting Lagrangian reads:
$${\cal L}=\frac{1}{2}\phi^{2}\left(-\partial^{2}+\frac{e^{2}}{\pi}M_{D}^{-1}\right)\phi+$$
\begin{equation} 
-\mu\cos\left(\sqrt{4\pi}\phi+
Q\alpha_{D}\right)+Q^{2}{\cal K}_{D},\label{eq:gen_bos_lagrangian}
\end{equation}
where $M_{D}$ and $\alpha_{D}$ can be seen as operators acting on $\phi$ and $K_{D}$ is a simple space-time function. 
After some algebra one can write $\alpha_{D}=2eF_{D}K$ and ${\cal K}_{D}=\frac{1}{2}\partial_{\mu}K\hat{F}_{D}\partial_{\mu}K$, where $F_{D}=G_{D}/\left(1+\frac{e^{2}}{\pi}G_{D}\right)$.
The unperturbed theory, i.e. the theory with no external charges, can be easily recovered by setting $Q=0$. 

Despite the non-locality, the above Lagrange is
still translational invariant in space and in time. The latter gives rise to the conservation of energy. 
The total energy can be computed through the energy-momentum tensor. This will be, in general, rather complicated. Nonetheless it is still possible to compute the difference of energies, since the more complicated terms cancel out (for the massive case one can also do it in first order in perturbation theory on the mass). 

In Appendix \ref{sec:EMtensor} the construction of the energy-momentum
tensor for theories with higher derivatives is revised.  In the present case there is an arbitrarily high number of derivatives. The energy is given
by the integral in space of the $T^{00}$ component of the energy-momentum
tensor, Eq. (\ref{eq:en-mom_tensor}). When quantized the
fields are promoted to operators and take $T^{00}$ in normal order so the total
energy is given by $E=\left\langle \int dx:T^{00}\left(x\right):\right\rangle $. 

The difference of energy as a result of introducing external
charges can be written as:
\begin{equation}
\Delta E=\left\langle \int dx:T_{Q}^{00}\left(x\right)-T_{Q=0}^{00}\left(x\right):,\right\rangle 
\end{equation}
where it was denoted by $T_{Q}^{00}\left(x\right)$ the energy-momentum
tensor when external charges are also present according to Eq. (\ref{eq:gen_bos_lagrangian}). 

The analysis start in the massless case moving then to the small mass limit.

\subsection{Massless fermions}

For massless fermions $\mu =0$. This case is much simpler to analyse since the effect of external charges 
is only in the term $Q^{2}{\cal K}_{D}$, completely decoupled from the fields.
However complicated the energy momentum tensor obtained
from the system without external charges is, the same tensor results from the presence of external 
charges with an additional space-time function independent of the fields. In other words, 
the first term of the tensor in (\ref{eq:en-mom_tensor})
is not affected by the external charges while the second 
is just ``translated'' 
with no operator content [see $Q^{2}K_{D}$ in (\ref{eq:gen_bos_lagrangian})].
Therefore for the massless case one obtains $\Delta E_{m=0}=Q^{2}\int dx{\cal K}_{D}\left(t=0,x\right)$. Recall that ${\cal K}_{D}=\frac{1}{2}\partial_{\mu}K\hat{F}_{D}\partial_{\mu}K$. Each $\partial_{\mu} K$ encodes two Dirac deltas 
corresponding to two different external charges as described above. Therefore in this expression there are 
included interactions between the charges, corresponding to pick the Dirac deltas at different points, 
and ``self-interactions'', corresponding to pick the same Dirac delta in both $K$'s. 
The latter ones are independent of $L$ and therefore do not account for actual interaction between different charges. 
In the following they are then neglected. By performing the implicit integrals on the definition of ${\cal K}_{D}$ and 
making use of the fact that $\partial_{\mu} K$ is time-independent, the energy can be then written as

\begin{equation}
\Delta E_{m=0}=Q^{2}\int\frac{dk_{1}}{2\pi}F_{D}\left(k_{0}=0,k_{1}\right)\exp\left(ik_{1}L\right)
\label{En_massless}
\end{equation}
where $F_{D}\left(k_{0},k_{1}\right)$ are the Fourier components of $F_{D}$.

In the following these integrals are computed explicitly for $D=1,2,3$.

\subsubsection{Massless $D=1$}

This case corresponds to have $\hat{G_1}=1/-\partial^2$ and $F_1\left(0,k\right)=1/\left(e^2/\pi+k^2\right)$. 
The integral results in:

\begin{equation}
\Delta E_{m=0}=\frac{\sqrt{\pi}Q^{2}}{4e}\exp \left(-\frac{eL}{\sqrt{\pi}}\right).
\end{equation}

Without the presence of the fermion field (or turning off the coupling $e=0$), the resulting energy would exhibit 
a linear growth with the distance. The exponential decay present here is a result of pair production that screens the external charges. This shows explicitly the charge screening known for the massless Schwinger model. 
As discussed in Section \ref{sec:robustness}, the same happens in presence of the Thirring interactions between fermions.

\subsubsection{Massless $D=2$}

The function $F_2$ is given by 
$F_2\left(0,k\right)=1/\left(e^2/\pi+2 \left|k\right|\right)$. 
Again the integral 
can be performed explicitly:

$$
\Delta E_{D=2}=\frac{Q^2}{2\pi} \Bigg[\frac{\pi}{2}\sin\left(\frac{e^2L}{2\pi}\right)
$$
\begin{equation}
-\cos\left(\frac{e^2L}{2\pi}\right) \mathrm{Ci}\left(\frac{e^2L}{2\pi}\right)
-\sin\left(\frac{e^2L}{2\pi}\right)\mathrm{Si}\left(\frac{e^2L}{2\pi}\right)\Bigg]
\end{equation}
The functions $\mathrm{Ci}$ (cosine integral) and 
$\mathrm{Si}$ (sine integral) are respectively given by 
$\mathrm{Ci}(x)=-\overset{+\infty}{\underset{-x}{\int}}dt\cos t/t$ and 
$\mathrm{Si}(x)=\overset{x}{\underset{0}{\int}}dt\sin t/t$. 
In the limit of $L\rightarrow\infty$ the cosine integral goes to zero 
and the sine integral converges to $\pi/2$. As a result also here 
the energy goes to zero (as $1/L^2$) as the distance $L$ 
increases despite the pure gauge theory is exhibiting a logarithm 
increase of the energy with the distance.

\subsubsection{Massless $D=3$}

For the three-dimensional case one has to introduce an UV cut-off $\Lambda$ 
in order to regularize the integral over the extra dimensions 
where the gauge field lives. The resulting function $F$ will be dependent 
on this cut-off: it is found $$F\left(0,k\right)=\frac{\log\left(1+\left(\Lambda/k\right)^2\right)}{4\pi+\frac{e^2}{\pi}\log\left(1+\left(\Lambda/k\right)^2\right)}.$$ The integral (\ref{En_massless}) requires a careful study. ù
Within a change of variables it can be written as:

\begin{equation}
\frac{Q^{2}}{4\pi L}\tilde{\Lambda}\overset{+\infty}{\underset{0}{\int}}\frac{dq}{2\pi}\frac{\log\left(1+q^{-2}\right)}{1+\left(e^{2}/4\pi\right)\log\left(1+q^{-2}\right)}\cos\tilde{\Lambda}q
\label{eq:massless3D}
\end{equation}

The dependence on the distance is now isolated in the prefactor $1/L$, 
since the remaining was absorbed into the cut-off $\tilde{\Lambda}=L\Lambda$. 
In this expression the screening due to pair creation is evident: 
setting $e=0$ the integral (\ref{eq:massless3D}) simply gives $\tilde{\Lambda}^{-1}$ (in the large cut-off limit) and what it remains is the expected 
Coulomb energy: $Q^{2}/4\pi L$. When $e$ acquires a finite value, 
one is coupling the gauge fields to the fermion fields and the 
pair production starts. This is made explicit in the integral since this extra 
positive term in the denominator will decrease the absolute value of the integrand. 

It is now shown that for any finite charge $e$ total screening occurs and actually $\Delta E_{D=3}=0$ in the large cut-off limit.
The integral can be broken into pieces: 
$\int_{0}^{+\infty}=\sum_{n}\int_{2\pi n/\Lambda}^{2\pi\left(n+1\right)/\Lambda}$. 
Let us denote the 
non-oscillatory part by 
$f\left(q\right)=\log\left(1+q^{-2}\right)/
\left(1+\left(e^{2}/4\pi\right)\log\left(1+q^{-2}\right)\right)$. 
Since this function is not singular except for $q=0$, most of the 
integrals to be computed vanish in the large cut-off limit. 
This can be seen by integrating by parts which will bring powers of 
$\tilde{\Lambda}$ to the denominator. At lowest order, 
if the functions $f$ has finite derivatives, the leading term goes 
like $\tilde{\Lambda}^{-3}$. 
After the above procedure, the only part that remains is the case $n=0$. This is treated by observing that 
$f$ is strictly decreasing in the interval of integration. Therefore

\begin{equation}
\tilde{\Lambda}\overset{2\pi/\tilde{\Lambda}}{\underset{0}{\int}}\frac{dq}{2\pi}f\left(q\right)\cos\tilde{\Lambda}q\leq2\left[\frac{4\pi}{e^{2}}-f\left(\frac{2\pi}{\tilde{\Lambda}}\right)\right]
\end{equation}
as one can see by replacing the value of the function $f$ by its maximum value 
in the interval ($4\pi/e^2$) when the cosine is positive and by 
its minimum value when the cosine is negative. Since the function is 
continuous one can make $f\left(2\pi/\tilde{\Lambda}\right)$ as close 
as desired to $4\pi/e^2$ by increasing $\tilde{\Lambda}$, 
therefore the bound goes to zero. Since the integral is positive this shows that the energy goes to zero. One can finally write:

\begin{equation}
\Delta E_{D=3}=
\begin{cases} 
      \frac{Q^{2}}{4\pi L} & e=0 \\
      0 & e\neq 0 
\end{cases}
\end{equation}

When the coupling between the gauge fields and the fermions is turned on, 
the fermionic fields react to the presence of external charges 
by starting pair production. Remarkably they are able to completely 
screen the external charges. This suggests that when the gauge field is in $3+1$ 
dimensions the fermions become more effective at screening 
external charges than at $2+1$ or even $1+1$. For the latter case 
the energy decreases exponentially with the distance while here 
it is zero for any distance.

\subsection{The massive case}

The massive case is more complicated and the effect of the 
external charges is no longer decoupled from the fields. 
Without the external charges, the interaction is still local 
and all the non-locality
is on the kinetic term. When the external charges are introduced, 
the non-locality is carried over to the interaction via $\alpha_{D}$.
This means that the insertion of external charges will modify both terms present in (\ref{eq:en-mom_tensor}). 

Let us start when the system has no external charges. By performing 
first order perturbation theory in the mass of the fermionic theory, 
the ground-state has the structure: $\left|\Omega_{0}\right\rangle =\left|0\right\rangle +\mu\left|1\right\rangle$.
The state $\left|0\right\rangle $ is the vacuum of the massless theory, 
which is still a quadratic theory. The normal ordering
is taken with respect to this state. When going to the system with
external charges, even though the quadratic term is modified, the modification  
is of order $\mu$ so that the ground state of such theory is,
at the lowest order in perturbation theory, 
given by $\left|\Omega_{Q}\right\rangle =\left|0\right\rangle +
\mu\left|1^{\prime}\right\rangle$, 
where $\left|0\right\rangle $ is the same vacuum state of the massless
theory and the first order correction was modified 
due to the presence of external charges. The normal ordering is then
taken with respect to the same state in both theories. 

One can now analyse the energy-momentum tensor 
(\ref{eq:en-mom_tensor}). 
With no external charges $T^{00}=T_{0}-\mu\cos\left(\sqrt{4\pi}\phi\right)$,
where all the contributions independent of $\mu$ were inserted in $T_{0}$.
In the presence of external charges, this is modified 
to $T^{00}=T_{0}+\mu\tilde{T}_{0}-\mu\cos\left(\sqrt{4\pi}\phi+
Q\alpha_{D}\right)+Q^{2}K_{D}$ 
where $\mu\tilde{T}_{0}$ is the order-$\mu$ term obtained from the
first part of (\ref{eq:en-mom_tensor}). Explicitly one has:
$$
\tilde{T}_{0}=\underset{n=0}{\overset{+\infty}{\sum}}\underset{i=0}{\overset{n}{\sum}}\left(-1\right)^{i}\partial_{\mu_{1}}\ldots\partial_{\mu_{i}}\frac{\partial\cos\left(\sqrt{4\pi}\phi+Q\alpha_{D}\right)}{\partial\left(\partial_{0}\partial_{\mu_{1}}\ldots\partial_{\mu_{n}}\phi\right)} \, \cdot
$$
\begin{equation}
\cdot \, \partial_{\mu_{i+1}}\ldots\partial_{\mu_{n}}\partial^{0}\phi\label{eq:modified_kin}
\end{equation}
The presence of this term is due to the fact that
the non-locality is carried over to the interacting part, proportional
to $\mu$, by the presence of external charges. As a result one has
at first order of perturbation theory:
$$
\Delta E_{m}=\Delta E_{m=0}+\mu\left\langle 0\right| 
\int dx \, :\bigg[ \tilde{T}_{0}
$$
\begin{equation}
+\cos\left(\sqrt{4\pi}\phi\right)-\cos\left(\sqrt{4\pi}\phi+Q\alpha_{D}\right):\bigg] \left|0\right\rangle.
\end{equation}

Due to the normal ordering, the only non-vanishing term from 
the Taylor expansion of the first cosine is $1$. All the others 
average to zero in the ground state. The same
kind of argument holds for $\cos\left(\sqrt{4\pi}\phi+Q\alpha_{D}\right)$, 
where only $\cos\left(Q\alpha_{D}\right)$ is non vanishing. Finally, it is
observed that, by construction, any term of $\tilde{T}_{0}$ has always
at least one $\phi$, as it is clear from \ref{eq:modified_kin}. 
Therefore it averages to zero in the ground state in presence of 
normal ordering. The result is then:
\begin{equation}
\Delta E_{m}=\Delta E_{m=0}+\mu\overset{+\infty}{\underset{-\infty}{\int}} dx\left[1-\cos\left(Q\alpha_{D}\right)\right].\label{eq:energy_general}
\end{equation}

Intuitively, from the above expression one expects a finite string tension when
$Q\alpha_{D}$ is "mostly" non multiple of $2\pi$ between $-L/2$ and
$L/2$ and "mostly" multiple of $2\pi$ outside this interval. This,
as it will be seen explicitly in the following, is what happens for the
derived $\alpha_{D}$ in the different dimensions. From the definition of 
$\alpha_D$, and following the same path used for ${\cal K}_D$ when deriving 
(\ref{En_massless}), one can write: 
\begin{equation}
\alpha_{D}\left(0,x\right)=8e\overset{+\infty}{\underset{0}{\int}}\frac{dk}{2\pi}\frac{F\left(k_{0}=0,k\right)\sin\left(kL/2\right)\cos\left(kx\right)}{k}
\end{equation}

Again the different dimensions are considered separately 
in order to compute the increment to the energy due to the presence
of the mass. From this point on the notation 
$\alpha_{D}\left(x\right)\equiv\alpha_{D}\left(0,x\right)$ is adopted.

\subsubsection{Massive $D=1$}

This is the well known case studied in detail in the literature 
\cite{zinnjustin}. In this Section the computation to retrieve 
the expected results in the present formalism is performed, serving also to set the scene and the notation for the subsequent computations in $D=2$ and $D=3$. One has

\begin{equation}
\alpha_{1}\left(x\right)=8e\overset{+\infty}{\underset{0}{\int}}\frac{dk}{2\pi}\frac{\sin\left(kL/2\right)\cos\left(kx\right)}{k\left(k^{2}+\frac{e^{2}}{\pi}\right)}
\end{equation}
This integral can be calculated explicitly giving:
$$
\alpha_{1}\left(x\right)=\frac{\pi}{e}\left[\mathrm{sign}\left(L-2x\right)\left(1-\cosh\left(\frac{e}{2\sqrt{\pi}}\left(L-2x\right)\right)\right.\right.
$$
$$
\left.\left.+\sinh\left(\frac{e}{2\sqrt{\pi}}\left|L-2x\right|\right)\right)+\mathrm{sign}\left(L+2x\right)\left(1\right.\right.
$$
\begin{equation}
\left.\left.-\cosh\left(\frac{e}{2\sqrt{\pi}}\left(L+2x\right)\right)+\sinh\left(\frac{e}{2\sqrt{\pi}}\left|L+2x\right|\right)\right)\right]
\label{alpha_1}
\end{equation}
It turns out that, in order to compute the string tension, it is enough 
to work out the limits $\left|x\right|\ll L/2$ and $\left|x\right|\gg L/2$, 
as it is shown below. By inspecting directly the function (\ref{alpha_1}) one finds:
\begin{equation}
\alpha_{1}\left(x\right)=\left\{ \begin{array}{cc}
\frac{2\pi}{e} & \mathrm{\ if}\ \left|x\right|\ll L/2\\
0 & \mathrm{\ if}\ \left|x\right|\gg L/2
\end{array}\right.\label{eq:assymptotic_alpha1}
\end{equation}
This is enough to compute the string tension
from (\ref{eq:energy_general}) even without computing exactly the
integral (\ref{eq:energy_general}) itself. 
Indeed, making use of the fact that the integrand is symmetric
over $x\rightarrow-x$, the integral can be broken in three parts: 
$$\overset{+\infty}{\underset{0}{\int}}=\overset{L/2-x_{0}}{\underset{0}{\int}}+\overset{L/2+x_{0}}{\underset{L/2-x_{0}}{\int}}+\overset{+\infty}{\underset{L/2+x_{0}}{\int}}.$$
The value of $x_{0}$ is fixed to 
guarantee that $\exp\left(-\frac{e}{\sqrt{\pi}}\left(L/2-x_{0}\right)\right)\ll1$. Within this limit one can compute the first and the third integrals 
using the asymptotic expressions in (\ref{eq:assymptotic_alpha1}) obtaining:
$$\Delta E_{m}=\Delta E_{m=0}+\left[1-\cos\left(\frac{2\pi Q}{e}\right)\right]\left(L-2x_{0}\right)+$$
\begin{equation}
+2\overset{L/2+x_{0}}{\underset{L/2-x_{0}}{\int}}dx\left\{1-\cos\left[Q\alpha_{1}\left(x\right)\right]\right\},
\end{equation}
where one can explicitly see the linear growth in $L$. Note that 
$x_{0}$ can be chosen independent of $L$ for large enough values of $L$, 
and the corresponding term $\propto x_0$ in $\Delta E_m$ actually does
not grow with $L$. This reflects the contribution arising for the two regions 
close to the charges which is independent of their distance (if large enough). 
Furthermore, the remaining integral is bounded
by values independent of $L$. Explicitly, substituting the cosine
by $-1$ one has an upper bound of $4x_{0}$ and substituting the cosine
by $1$ one has a lower bound of $0$. 

One can therefore conclude that
the linear behaviour in $L$ is exclusive of the term $\propto L$ and one
can finally write:
\begin{equation}
\Delta E_{m}=\Delta E_{m=0}+\mu\left[1-\cos\left(\frac{2\pi Q}{e}\right)\right]L+\ldots
\end{equation}
where the dots indicate some bounded dependence on $L$. The string
tension reads explicitly:
\begin{equation}
\sigma_{1}=\mu\left[1-\cos\left(\frac{2\pi Q}{e}\right)\right]
\end{equation}
This is the well known result \cite{zinnjustin}, reviewed also in 
Appendix \ref{sec:massive_tension}, obtained here by a
careful analysis of the energy excess due to the presence of external charges. 
For higher dimensions the same procedure shall be followed in the next Sections.

\subsubsection{Massive $D=2$}

For this case one has:

\begin{equation}
\alpha_{2}\left(x\right)=4e\overset{+\infty}{\underset{0}{\int}}\frac{dk}{2\pi}\frac{\sin\left(kL/2\right)\cos\left(kL\right)}{k\left(k+\frac{e^{2}}{2\pi}\right)}
\end{equation}
which again can be computed explicitly. For brevity it was denoted
$X^{\pm}=L\pm2x$ and for $x\in\left[-L/2,L/2\right]$ one obtains:

$$
\alpha_{2}\left(x\right)=\frac{2}{e}\left[\pi-\frac{\pi}{2}\cos\left(\frac{e^{2}X^{-}}{4\pi}\right)-\frac{\pi}{2}\cos\left(\frac{e^{2}X^{+}}{4\pi}\right)\right.
$$
$$
\left.-\text{Ci}\left(\frac{e^{2}X^{+}}{4\pi}\right)\sin\left(\frac{e^{2}X^{+}}{4\pi}\right)-\text{Ci}\left(\frac{e^{2}X^{-}}{4\pi}\right)\sin\left(\frac{e^{2}X^{-}}{4\pi}\right)\right.
$$
\begin{equation}
\left.+\text{Si}\left(\frac{e^{2}X^{+}}{4\pi}\right)\cos\left(\frac{e^{2}X^{+}}{4\pi}\right)+\text{Si}\left(\frac{e^{2}X^{-}}{4\pi}\right)\cos\left(\frac{e^{2}X^{-}}{4\pi}\right)\right].
\end{equation}
For $\left|x\right|>L/2$:

$$
\alpha_{2}\left(x\right)=\frac{2}{e}\left[\frac{\pi}{2}\cos\left(\frac{e^{2}X^{-}}{4\pi}\right)-\frac{\pi}{2}\cos\left(\frac{e^{2}X^{+}}{4\pi}\right)\right.
$$
$$
\left.+\frac{\pi}{2}\sin\left(\frac{e^{2}x}{2\pi}\right)-\text{Ci}\left(\frac{e^{2}X^{+}}{4\pi}\right)\sin\left(\frac{e^{2}X^{+}}{4\pi}\right)\right.
$$
\begin{equation}
\left.+\text{Si}\left(\frac{e^{2}X^{-}}{4\pi}\right)\cos\left(\frac{e^{2}X^{-}}{4\pi}\right)+\text{Si}\left(\frac{e^{2}X^{+}}{4\pi}\right)\cos\left(\frac{e^{2}X^{+}}{4\pi}\right)\right].
\end{equation}
The same procedure from before is followed, studying the limits
of $\left|x\right|\ll L/2$ and $\left|x\right|\gg L/2$. The result
obtained is actually the same here. For $\left|x\right|\ll L/2$ this is seen by noticing that inside the cosine and sine integrals
one can replace $X^{\pm}$ by $L$ and take the large $L$ limit. Then
one can replace the cosine integral by zero and the sine integral by
$\pi/2$. For the second case $X^{\pm}$ is replaced by $\pm2x$
inside the cosine and sine integrals 
[note also that $\text{Si}\left(-y\right)=-\text{Si}\left(y\right)$].
Then one finds the same result as for (\ref{eq:assymptotic_alpha1})
and the results for the $1+1$ case translates directly to $2+1$. In particular the string tension is the same at this order
in perturbation theory in the mass:

\begin{equation}
\sigma_{2}=\mu\left[1-\cos\left(\frac{2\pi Q}{e}\right)\right]=\sigma_{1}
\end{equation}
Even though te presence of confinement in itself is not 
a surprise for this case, it is interesting
to observe that the the resulting string tension, at this order in perturbation
theory, is independent of the gauge fields living 
in $1+1$ or $2+1$ dimensions.

\subsubsection{Massive $D=3$}

Finally the case in which the gauge field lives in 
$3+1$ dimensions is considered. The function $\alpha$ reads:
\begin{equation}
\alpha_{3}\left(x\right)=\frac{2e}{\pi}\overset{+\infty}{\underset{0}{\int}}\frac{dk}{2\pi}\frac{\log\left(\frac{\Lambda^{2}+k^{2}}{k^{2}}\right)\sin\left(kL/2\right)\cos\left(kx\right)}{k\left(1+\frac{e^{2}}{4\pi^{2}}\log\left(\frac{\Lambda^{2}+k^{2}}{k^{2}}\right)\right)}
\end{equation}
The same kind of analysis, that was done for the other two cases, can be followed here.
This consists in looking at the two limits where $x$ is much smaller or much larger than $L$.
The three main steps consist on writing  $2\sin\left(kL/2\right)\cos\left(kx\right)=\sin\left(k\left(L/2+x\right)\right)+\sin\left(k\left(L/2-x\right)\right)$, breaking the integral in two contributions and performing the substitution
$q=k/\Lambda$. A factor is absorbed in the cut-off 
$\tilde{\Lambda}=\Lambda\left|L/2\pm x\right|$ choosing
$\pm$ depending on the argument of the sign in each piece. Since the limit of interest is where $x$ is far away from
$L/2$, this rescaling is well defined and the limit 
$\tilde{\Lambda}\rightarrow+\infty$
still makes sense. The integral reads:
$$
\alpha_{3}\left(x\right)=\frac{e}{\pi} 
\left[\mathrm{sign}\left(L/2+x\right)+\mathrm{sign}\left(L/2-x\right)\right] 
\cdot$$
\begin{equation}
\cdot \overset{+\infty}{\underset{0}{\int}}\frac{dq}{2\pi}\frac{\log\left(1+q^{-2}\right)\sin\left(\tilde{\Lambda}q\right)}{q\left(1+\frac{e^{2}}{4\pi^{2}}\log\left(1+q^{-2}\right)\right)}
\end{equation}
Each of the ``$\mathrm{sign}$'' terms comes respectively from
$\sin\left(k\left(L/2+x\right)\right)$ and 
$\sin\left(k\left(L/2-x\right)\right)$ factors 
in order to take care of the correct sign. One immediately sees 
that if the signs of $L/2\pm x$
are different, as they are in one of the relevant cases $\left|x\right|\gg L/2$,
$\alpha_{3}$ is zero. 

For the other case of $\left|x\right|\ll L/2$
the integral bears similarities to (\ref{eq:massless3D}) 
and part of the approach
can be followed. Namely, one can divide the integral
in small pieces $\int_{0}^{+\infty}=\sum_{n}\int_{2\pi n/\tilde{\Lambda}}^{2\pi\left(n+1\right)/\tilde{\Lambda}}$
and observe that several of them converge to zero as the limit
of large cut-off is taken. This is due to the rapid oscillation of the sine (or cosine as in Eq. (\ref{eq:massless3D})]. Then the only remaining part
is:

\begin{equation}
\alpha_{3}\left(x\right)=\frac{2e}{\pi}\overset{2\pi/\tilde{\Lambda}}{\underset{0}{\int}}\frac{dq}{2\pi}\frac{\log\left(1+q^{-2}\right)\sin\left(\tilde{\Lambda}q\right)}{q\left(1+\frac{e^{2}}{4\pi^{2}}\log\left(1+q^{-2}\right)\right)},\ \left|x\right|<L/2.
\end{equation}
In (\ref{eq:massless3D}) the part analogous to this last piece was also zero as long as $e$ was finite. Now this is no longer correct due to the $1/q$ factor which
picks a large contribution near $q=0$. To see this explicitly, one
takes the leading order of $\log\left(1+q^{-2}\right)/\left(1+\frac{e^{2}}{4\pi^{2}}\log\left(1+q^{-2}\right)\right)$
for small $q$ which is simply $\frac{4\pi^{2}}{e^{2}}$. 
Then the result is independent of the
cut-off $\tilde{\Lambda}$:
\begin{equation}
\alpha_{3}\left(x\right)=\frac{4}{e}\overset{2\pi/\tilde{\Lambda}}{\underset{0}{\int}}dq\frac{\sin\left(\tilde{\Lambda}q\right)}{q}=\frac{4\mathrm{Si}\left(2\pi\right)}{e},\ \left|x\right|<L/2.
\end{equation}
Summing up then the results:
\begin{equation}
\alpha_{3}\left(x\right)=\left\{ \begin{array}{cc}
\frac{4\mathrm{Si}\left(2\pi\right)}{e} & \mathrm{\ if}\ \left|x\right|<L/2\\
0 & \mathrm{\ if}\ \left|x\right|>L/2
\end{array}\right.
\end{equation}
which again corresponds to the expected behaviour for a confined phase.
The string tension is given by: 
\begin{equation}
\sigma_{3}=\mu\left[1-\cos\left(\frac{4\mathrm{Si}\left(2\pi\right)Q}{e}\right)\right]
\label{results_D3}
\end{equation}
which is finite in general. It is interesting to observe that even though
$\sigma_{1}=\sigma_{2}$, nevertheless they are still different from $\sigma_{3}$.
Furthermore, in the two previous cases if the external charge
$Q$ is a multiple integer of $e$ the string tension vanishes. For $D=3$ it is
no longer the case. The string tension remains finite when $Q$ is a multiple
integer of $e$ with the factor of $2\pi$ in the argument of cosine 
replaced in (\ref{results_D3}) 
by $4\mathrm{Si}\left(2\pi\right)\simeq5.67<2\pi$. 
Total screening is obtained instead for 
$Q=\frac{\pi}{2\mathrm{Si}\left(2\pi\right)}e$.

\section{Conclusions}
\label{conclusions}

In this work, the robustness of the confined phase 
for $1+1$ fermions was studied by determining the string tension between two probe static 
charges. In the first part of the paper, the 
robustness of the confinement properties when a Thirring 
interaction term is added to the Schwinger model was investigates, as well as the effect of a topological $\theta$-term.
Through bosonization, it was shown that the
known results for both models (i.e., when different types of interactions are separately considered) still hold. 
The theory only makes sense when the Thirring coupling is $g>-\pi$
(as in the Thirring model) and most importantly, the system only
deconfines for $\theta=\pm\pi$ (as in the Schwinger model). Through
an Hubbard-Stratonovich transformation, one observes that this model
can be regarded as a fermionic field interacting with a massless gauge
field which in turn interacts with a `{\it massive} gauge field. It
is possible that general interactions of this form may break confinement. 
This can be an interesting problem to address in the future since, 
in principle, such interactions may be experimentally available in the context of quantum simulations. 
However, the terms obtained from the Thirring interaction 
induce cancellations which would
not appear under a general coupling between the gauge fields. 
The Thirring parameter does not allow to vary interaction between 
the bosonic fields but only the mass of one of them. This result shows that, 
with respect to confinement, a 
possible nearest-neighbour density-density interaction
plays no role and therefore the phase is stable. 
In the case considered here the static charges coupled directly 
to the gauge fields (through $-iQA_{\mu}j_{\mathrm{ext}}^{\mu}$), 
but not directly to the fermion current, 
even in the presence of the Thirring term. Such term would take the 
form $\frac{1}{2}g_{\mathrm{ext}}\bar{\psi}\gamma_\mu\psi j_{\mathrm{ext}}^{\mu}$ 
and, therefore, has the appearance of an external field. 
The study of such systems is beyond the scope of the present work, but 
it may also be theoretically and experimentally relevant. 
Estimates of the parameters for a possible experimental implementation 
with ultracold atoms in optical lattices have been also presented. 

The possibility of higher dimension of the gauge fields 
while the fermions remain in $1+1$ dimensions was also considered. It was found that in the massless case, as for the Schwinger model, there is a strong screening when static charges are introduced on the system. In the Schwinger case the linear growth of the energy with the distance is replaced by an exponential decay. In the case of a $2+1$ dimensional gauge field the logarithm 
is replaced by oscillatory functions (which go to zero as power laws). 
Finally in $3+1$ dimensions the $1/L$ decay is replaced by zero: 
external charges are completely screened. When a small mass is considered, 
a linear growth of the energy with the distance is observed and 
therefore a finite string tension is obtained for all 
$1+1$, $2+1$ or $3+1$ dimensions. Furthermore, at this order in perturbation 
theory, the string tension 
is the same for the first two cases and smaller for the latter 
one: $\sigma_1=\sigma_2>\sigma_3$. 
The last inequality only holds for 
small enough external charges (notice anyway that 
the string tensions are periodic functions of the external charges).

This result is somewhat counter-intuitive since the confinement in 
the Schwinger model is usually attributed to the fact that the 
Gauss law in $1+1$ dimensions impose a constant electric field 
(rather than $1/r^2$ of the $3+1$ system). Our results suggest that 
this feature is not necessary to obtain confinement and, 
instead, it is the dimensionality of the space-time available to the fermion 
fields that is rather dictating confinement. In order to better test 
this hypothesis it would be interesting to study how far can one extend the
space-time allowed for the fermion fields before leaving the confined phase 
(for gauge fields in $3+1$ dimensions for example). Given that 
it is known that when the fermion fields span the full $3+1$ dimensions 
the theory is deconfined (corresponding to regular QED), 
this point of transition does exist. Furthermore, with the 
advent of quantum simulation of gauge theories, one can hope that 
an experiment with tunable fermion dimensionality~\cite{Lamporesi:2010aa} could probe 
directly interesting phenomena like this transition.

We finally observe that the case of the gauge fields living in 
$3+1$ dimensions exhibits a quantitative difference with respect 
to the other two in which the gauge fields are defined in $1+1$ and $2+1$. 
It would be interesting to understand in detail how 
the models differ. A quantity of interest would be the 
expected fermion distribution in the presence of external charges. 
In the $3+1$ case the matter-gauge system creates, dynamically, 
a linear growth of the energy from a static $1/r^2$ energy interaction. T
he way this happens is expected to be quantitatively different from the case 
of $1+1$ dimensions, where the static energy interaction is already linear.

{\it Acknoledgements:} Discussions with L. Lepori, P. Minkowski, A. A. Nersesyan, A. M. Tsvelik and U.-J. Wiese  are gratefully acknowledged. MD work is partly supported by the ERC Starting grant AGEnTh (758329).

\bibliography{ST}

\bibliographystyle{apsrev4-1}

\appendix

\section{Particle spectrum for the massless Schwinger-Thirring model}
\label{sec:2pntfunctions}

As in the case of the Schwinger model \cite{zinnjustin}, the
divergences of the correlation functions $\bar{\psi}\psi$ and 
$\bar{\psi}\gamma_{S}\psi$ are of the form 
$\left(1+\frac{g}{\pi}\right)p^{2}=-n^{2}e^{2}/\pi$. 
The case $n=1$ is the only simple
pole and comes from the pseudoscalar two-point function. 
The connection between the four fermion functions and the propagator 
can be studied using the approach 
for the Schwinger model described in \cite{zinnjustin} 
in the presence of the Thirring term. 

From the bosonization procedure described in Section \ref{sec:robustness}, 
one has $\bar{\psi}^{\prime}\psi^{\prime}\propto\cos\left(\sqrt{4\pi}\vartheta^{\prime}\right)$
and $\bar{\psi}^{\prime}\gamma_{S}\psi^{\prime}\propto i\sin\left(\sqrt{4\pi}\vartheta^{\prime}\right)$, 
where the $\psi^{\prime}$ and $\vartheta^{\prime}$ are the fermionic
and bosonic intermediate fields used in the calculation. The relations 
with the initial fermionic bosonic variables are 
$\bar{\psi}\psi\cos2e\mbox{\ensuremath{\varphi}}-i\bar{\psi}\gamma_{S}\psi\sin2e\mbox{\ensuremath{\varphi}}\propto\cos\left(\sqrt{4\pi}\vartheta-2e\varphi\right)$
and $\bar{\psi}\gamma_{S}\psi\cos2e\mbox{\ensuremath{\varphi}}-i\bar{\psi}\psi\sin2e\mbox{\ensuremath{\varphi}}\propto i\sin\left(\sqrt{4\pi}\vartheta-2e\varphi\right)$.
From these relations one obtains 
$\bar{\psi}\psi\propto\cos\left(\sqrt{4\pi}\vartheta\right)$
and $\bar{\psi}\gamma_{S}\psi\propto\sin\left(\sqrt{4\pi}\vartheta\right)$.
Therefore the relations between the initial fermionic and final bosonic
fields are the same of the free theory: 

$$
\left\langle \bar{\psi}\left(x\right)\psi\left(x\right)\bar{\psi}\left(0\right)\psi\left(0\right)\right\rangle =\left\langle \bar{\psi}\psi\right\rangle \cosh\left(4\pi\Delta\left(x\right)\right)
$$
\begin{equation}
\left\langle \bar{\psi}\left(x\right)\gamma_{s}\psi\left(x\right)\bar{\psi}\left(0\right)\gamma_{s}\psi\left(0\right)\right\rangle =\left\langle \bar{\psi}\gamma_{s}\psi\right\rangle \sinh\left(4\pi\Delta\left(x\right)\right).
\label{eq:fermion4function}
\end{equation}
The singularities can be determined by expanding the $\cosh$ and the
$\sinh$ in power series and analysing them term by term. 
Let us consider the term of order $n$ which corresponds to $\cosh$ if even
or $\sinh$ if odd. 
By doing the Fourier transform of (\ref{eq:propagator}), 
exponentiating it and then Fourier transforming again, 
one can compute the Fourier transform of (\ref{eq:fermion4function}) 
in terms of the momentum $p$. The result is given by:
$$
\int\frac{d^{2}q_{1}}{\left(2\pi\right)^{2}}\cdots\frac{d^{2}q_{n}}{\left(2\pi\right)^{2}}\frac{1}{\left(1+\frac{g}{\pi}\right)q_{1}^{2}+\frac{e^{2}}{\pi}} 
$$
\begin{equation}
\cdots\frac{1}{\left(1+\frac{g}{\pi}\right)q_{n}^{2}+\frac{e^{2}}{\pi}}\delta\left(p-q_{1}-\ldots-q_{n}\right)
\end{equation}
The integration of one of the variables, say $q_{n}$, can be performed
using the Dirac delta. The $n-1$ integrations of the zero-th
component of $q_{i}$ can be then carried out putting them on-shell. This
results in:
$$
\int\frac{d\left(q_{1}\right)_{1}}{2\pi\sqrt{2E_{1}}}\cdots\frac{d\left(q_{n-1}\right)_{n-1}}{2\pi\sqrt{2E_{n-1}}} \cdot
$$

\begin{equation}
\cdot \frac{1}{\left(1+\frac{g}{\pi}\right)\left(p-q_{1}-\ldots-q_{n-1}\right)^{2}+\frac{e^{2}}{\pi}} \bigg|_{\begin{array}{l}
q_{i=1,\ldots,n-1}\\ 
\mathrm{on-shell}
\end{array}},
\label{eq_app}
\end{equation}
where it was abbreviated $E_{i}=\sqrt{q_{i}^{2}+m^{2}}$. Using the notation 
$Q=q_{1}+\ldots+q_{n-1}$, one can write the denominator of 
(\ref{eq_app}) in the form
$\lambda\left(p-q\right)^{2}+m^{2}$. The momenta part can be written
as $\left(p-q\right)^{2}\rightarrow p_{0}^{2}-2p_{0}Q_{0}+Q^{2}$, 
where it was used that it is possible to eliminate the dependence on the
spatial component of $p$ by a suitable translation of the spatial
variable of integration. The poles then obey 
$\sqrt{\lambda}p_{0}=\sqrt{\lambda}Q_{0}+\sqrt{Q_{1}^{2}-m^{2}}$.
Because the particles $q_{i}$ are on-shell, the maximum value of
the total for momenta is $\lambda Q^{2}=-\left(n-1\right)^{2}m^{2}$
corresponding to the situation where all the $n-1$ particles are
at rest in a given frame and therefore $Q_{1}=0$. For this case one
finds a pole at $\lambda p_{0}^{2}=-n^{2}m^{2}$. By increasing the
total momentum of $Q$ one finds a branch cut along the axis starting
at $-n^{2}m^{2}$ corresponding to multiparticle states.
This holds for any $n>1$. For the special case $n=1$ there is an
isolated pole at $\lambda p_{0}^{2}=-m^{2}$ and therefore the theory
does not contain further states.

\section{String tension for the perturbative massive case}
\label{sec:massive_tension}

The massive case can be addressed pertubatively. 
The same procedure described in Section (\ref{sec:robustness}) 
to derive the bosonic Lagrangian can be repeated, 
now for a system with the presence
of two external charges. 

Since the external charges are placed at a finite
distance $L$, terms with an external current of the form $J_{0}^{\mathrm{ext}}=\delta\left(x-L/2\right)-\delta\left(x+L/2\right)$
and $J_{1}^{\mathrm{ext}}=0$ should be added to the Lagrangian. The
Thirring term will produce no extra contribution for the string tension
as $\left(J_{\mu}^{\mathrm{ext}}\right)^{2}$ contains no element
involving the two different charges together (it is purely local) and,
therefore, will give an $L$ independent contribution for the final
energy. The effect of the external charges enters through the coupling
with an external field $-iQJ_{\mu}^{\mathrm{ext}}A_{\mu}$, being $Q$
the absolute value of the external charges placed on the system. 
After the variable transformation, 
this coupling is transformed into 
$-iQJ_{\mu}^{\mathrm{ext}}\left(C_{\mu}-B_{\mu}\right)$.
As in \cite{coleman1975charge} the effect of the external charges is 
easily seen if
one writes $J_{\mu}^{\mathrm{ext}}=\varepsilon_{\mu\nu}\partial_{\nu}K$.
This term takes then the form 
$QK\left(\partial^{2}\varphi-\partial^{2}\varphi^{\prime}\right)$.
The function $K$ is mostly constant being $1$ for $\left|x\right|<L$
and $0$ for $\left|x\right|>L$. This extra term has the form of
the $\theta$-term with the difference that $K$ is actually 
space-dependent. Therefore, when one does the 
transformation $\vartheta\rightarrow\vartheta-\sqrt{\pi}QK/e$, 
there is a kinetic term corresponding to the points $\left|x\right|=L$.
Again such contribution is independent of $L$ and it is not important
to compute the string tension. Since when $K=0$ the contribution
for the energy from both systems is the same, the difference of energy
corresponds to take simply $K=1$ and multiply the energy density
by $L$. 

In lowest order in perturbation theory in the mass, the energy
corresponds simply to the expectation value of the cosine term. One then finds 
the known result for the Schwinger model \cite{zinnjustin} 
 
\begin{equation}
\sigma=-\mu\left[\cos\left(\theta-\frac{2\pi Q}{e}\right)-\cos\left(\theta\right)\right].
\end{equation}

\section{Details on parameters estimate}\label{sec:parameter_estimate}

This Appendix provides further details on the estimation of
the parameters of the lattice model. The hopping parameter
of species $\alpha$ between nearest-neighbour sites of the optical lattices is denoted by $t_{\alpha}^{\vec{r}^{'}\vec{r}^{''}}$ and the interaction parameter
between species $\alpha$ and $\beta$ (assumed site independent) is denoted by $U_{\alpha\beta}$ . One has

$$
t_{\alpha}=-\int d^{d}\vec{r} \Bigg( 
\frac{\hbar^{2}}{2m_{\alpha}}\nabla\phi_{\alpha,n} \left(\vec{r}\right)\cdot\nabla\phi_{\alpha,n'}\left(\vec{r}\right)
$$
\begin{equation}
+\phi_{\alpha,n}\left(\vec{r}\right)V_{\mathrm{ext}}\left(\vec{r}\right)\phi_{\alpha,n'}\left(\vec{r}\right) \Bigg) \label{eq:hb_kinnetic}
\end{equation}
and 
\begin{equation}
U_{\alpha\beta}=g_{\alpha\beta}\int d^{d}\vec{r}\phi_{\alpha,n}\left(\vec{r}\right)^{2}\phi_{\beta,n}\left(\vec{r}\right)^{2}\label{eq:hb_interaction}
\end{equation}
where $\phi_{\alpha,n}$ is the Wannier function for the $\alpha$-species 
centred in the site $n$ of the optical lattice \cite{Jaksch98,Trombettoni01}. 
Furthermore $g_{\alpha\beta}=\frac{4\pi\hbar^{2}a_{\alpha\beta}}{m_{\alpha\beta}}$
where $a_{\alpha\beta}$ is the scattering length between species
$\alpha$ and $\beta$ and $m_{\alpha\beta}$ the corresponding reduced mass. In
the following it is assumed the $\phi$'s to be Gaussian: 
$\phi_{\alpha,n}\left(\vec{r}\right)=C_{\alpha}\underset{j=1}{\overset{3}{\prod}}e^{-\left(r_{j}-r_{n,\alpha;j}\right)^{2}/2\sigma_{\alpha,j}^{2}}$, 
where $\vec{r}=(r_1,r_2,r_3)$ and $\vec{r}_{n,\alpha}$ is the position of the $n$-th minimum of the lattice for the species $\alpha$ (the constants $C_\alpha$ enforcing the normalization). 
These functions are characterized by the $\sigma_{\alpha,j}$ which may depend 
on the direction and that are fixed by energy
minimization. For the present case the shape of the function
in the directions $y$ and $z$ is fixed through a parameter $\sigma_{\perp\alpha}$
while the value of the longitudinal component, which is called simply
$\sigma_{\alpha}$, can be fixed variationally. The relevant potentials, 
that can be realized experimentally, take
a form that can be written as: 
\begin{equation}
V\left(x\right)=V_{0}\left[\sin\left(kx\right)^{2}+\lambda\sin\left(2kx+\alpha\right)^{2}\right].
\end{equation}
To simplify the subsequent computation a quartic polynomial
potential is considered. It is constructed by expanding
the expression above in powers of $kx$ and having two minima, 
one at $-a/2$ and other at $a/2$, with an offset between them of  $\Delta V_0$. 
In order to derive the potential one can require that its first derivative is of
the form $\left(A/a\right)\left(2x/a+1\right)\left(x/a-x_{0}/a\right)\left(2x/a-1\right)$
where $x_{0}$ corresponds to the position of the maxima between the
two minima. The potential depth will be proportional to $A$. By integrating one obtains the form of the potential and
an extra parameter $c$ as a constant of integration. This parameter
is fixed by requiring that the absolute minima, chosen arbitrarily
to be the one at $x=-a/2$, corresponds to zero energy. The potential depth, $V_0$, is the height of the barrier at $x_0$ (position of the maximum). With this definition one can replace $A$ by $V_0$ according to $A=192V_0/\left(3-2x_0/a\right)\left(1 +2x_0/a\right)^3$.
The offset between the minima is $V_0\Delta$ where $\Delta=32x_0/a\left(3-2x_0/a\right) \left(1 +2x_0/a\right)^3$.
The potential considered for, say, the boson species $1$ is
then:

$$
V_{B1}\left(x\right)=A_{B} \left(\frac{x^4}{4a^4}-\frac{x^3 x_{B}}{3a^4}-\frac{x^2}{8a^2}\right.
$$
\begin{equation}
\left.+\frac{x x_{0B}}{4a^2}+\frac{x_{0B}}{12}+\frac{1}{64}\right).
\end{equation}
while for the boson species $2$ is $V_{B2}\left(x\right)=V_{B1}\left(-x\right)$.
For the fermions the potential $V_{F}\left(x\right)$ has the same structure 
with an amplitude $A_{F0}$: 
$$
V_{0F}\left(x\right)=A_{F} \left(\frac{x^4}{4a^4}-\frac{x^3 x_{0F}}{3a^4}-\frac{x^2}{8a^2}\right.
$$
\begin{equation}
\left.+\frac{x x_{0F}}{4a^2}+\frac{x_{0F}}{12}+\frac{1}{64}\right).
\end{equation}
It is important to observe, however, that only 
bosons -- unlike the fermions -- should feel a double well potential 
according to the proposal. For this reason the polynomial
double well potential approximation for the fermions is not as good
as an approximation as it is for the bosons. Nonetheless, as one is
interested in the strong coupling regime of the model, this provide 
a reasonable approximation, as we verified.

The difference between the energies of the minima, $V_0\Delta$, should be small when compared to $V_0$ or, equivalently, $x_0/a$ should be small. Consequently, their influence on the parameter determination is small, which was checked explicitly. In what follows it will then be taken $x_0/a=\Delta=0$ avoiding unnecessary complicated formulas. This results in
$$
t_{\alpha}=\frac{\hbar^{2}}{2m_{\alpha}a^2}\left[\frac{1}{4}\left(\frac{a}{\sigma_{\alpha}}\right)^{4}-\frac{1}{2}\left(\frac{a}{\sigma_{\alpha}}\right)^{2}-\frac{1}{\sigma_\perp^2}\right]e^{-\frac{a^{2}}{4\sigma_{\alpha}^{2}}}
$$
\begin{equation}
-V_0\left[12\left(\frac{\sigma_{\alpha}}{a}\right)^{4}-4\left(\frac{\sigma_{\alpha}}{a}\right)^{2}+1\right]e^{-\frac{a^{2}}{4\sigma_{\alpha}^{2}}}\label{eq:hb_kinnetic-1}
\end{equation}
and 
\begin{equation}
U_{\alpha\beta}=\frac{g_{\alpha\beta}}{2\pi^{3/2}}\frac{1}{\sigma_{\perp}^{2}}\frac{1}{\sqrt{\sigma_{\alpha}^{2}+\sigma_{\beta}^{2}}}\label{eq:hb_interaction-1}
\end{equation}
For the considered scheme it is required 
that $U=U_{11}=U_{22}$ and $U_{12}=U_{1F}=U_{2F}=2U$.
The fine-tuning of this condition is not crucial as discussed
in the main text. With the parameters $\sigma_{\perp}$ and $\sigma$
fixed, one has to rely on the control of the scattering length
in order to fulfil this condition. 

Within the variational approach,
one computes the average energy per site and requires that $\sigma$
minimizes it. The problem of the different shape of the minima, also
referred in the main text, can be addressed as follows.
The total energy is given by:

$$
\varepsilon=\int d^{3}\vec{r}\underset{\vec{r}',\alpha}{\sum}n_{\alpha}\frac{\hbar^{2}}{2m_{\alpha}}\left|\nabla\phi_{\alpha\vec{r}'}\right|^{2}+n_{\alpha}V_{ext}\left|\phi_{\alpha\vec{r}'}\right|^{2} +$$
\begin{equation}
+\underset{\beta>\alpha}{\sum}n_{\alpha}n_{\beta}\frac{g_{\alpha\beta}}{2}\left|\phi_{\alpha\vec{r}'}\right|^{2}\left|\phi_{\beta\vec{r}'}\right|^{2}
\end{equation}
($n_\alpha$ is the number of atoms per well of the species $\alpha$). 
Due to the asymmetry, the total energy per site is different depending
on which minima one is referring $V_{ext}\left|\phi_{\alpha\vec{r}'}\right|^{2}$.
For the minima at $x=\pm a/2$ the result is:

$$
\frac{\varepsilon}{N}=\underset{\alpha}{\sum}n_{\alpha}\frac{\hbar^{2}a^{-2}}{2m_{\alpha}}\frac{1}{2}\left(\frac{a}{\sigma_{\alpha}}\right)^{2}+$$
$$+\underset{\alpha}{\sum}4n_{\alpha}V_{0}\left(\frac{\sigma_\alpha}{a}\right)^2 \left(
3\left(\frac{\sigma_{\alpha}}{a}\right)^{2}+2\right)$$
\begin{equation}
+\underset{\alpha,\beta>\alpha}{\sum}n_{\alpha}n_{\beta}\frac{g_{\alpha\beta}}{4\pi^{3/2}\sigma_{\perp}^{2}\sqrt{\sigma_{\alpha}^{2}+\sigma_{\beta}^{2}}}\Bigg],
\end{equation}
where it was already included the approximation that the spreading in
the perpendicular direction is the same for all species and characterized
by $\sigma_{\perp}$. Assuming that all masses are the same and the
offsets $\Delta_{\alpha}=0$, the only
parameters species-dependent are the densities $n_{\alpha}$ and the amplitudes $V_{0\alpha}$. The
asymmetry of the minima is present whenever $\Delta_{\alpha}\neq0$.
Therefore the problem of the asymmetry of the different Wannier function
is not present here.
It was checked that the obtained estimates do not depend very much on the 
$\Delta$ parameters if they are not too large. The
density for fermions is $n_{F}=1$ while for the other two species
of bosons is $n_{1,2}=S$. With this there are in total two parameters
to fix, $\sigma_{B}$ and $\sigma_{F}$, given two coupled equations 
$\partial\varepsilon/\partial\sigma_{F}=0$ and 
$\partial\varepsilon/\partial\sigma_{B}=0$.
The reference energy is denoted by $E_{\mathrm{ref}}=\hbar^2/2ma^2$ with
an assumed equal mass for all the species $m$. The following dimensionless
parameters are now introduced: $\tilde{V}_{0\alpha}=V_{0\alpha}/E_{\mathrm{ref}}$, 
$\tilde{\sigma}_{\alpha}=\sigma_{\alpha}/a$
and $\tilde{a}_{\alpha\beta}=a_{\alpha\beta}/a$. 
Regarding the scattering lengths, one is working on the assumption
that $a_{1F}=a_{2F}=a_{12}\equiv2a_{\mathrm{scatt}}$ and 
$a_{11}=a_{22}=a_{\mathrm{scatt}}.$
The two equations are then:

\begin{equation}
\begin{split}
\left\{ \begin{array}{c}
\tilde{\sigma}_{F}^{-3}-16\tilde{V}_{0F}\tilde{\sigma}_{F}\left(3\tilde{\sigma}_{F}^{2}+1\right)
+\frac{4S\left(\tilde{a}_{1F}+\tilde{a}_{2F}\right)\tilde{\sigma}_{F}}{\sqrt{\pi}\tilde{\sigma}_{\perp}^{2}\left(\tilde{\sigma}_{F}^{2}+\tilde{\sigma}_{B}^{2}\right)^{3/2}}=0\\
\\
\tilde{\sigma}_{B}^{-3}-16\tilde{V}_{0B}\tilde{\sigma}_{B}\left(3\tilde{\sigma}_{B}^{2}+1\right)\\
+\frac{1}{\sqrt{\pi}\tilde{\sigma}_{\perp}^{2}}\left[\frac{\left(\tilde{a}_{1F}+\tilde{a}_{2F}\right)\tilde{\sigma}_{B}}{\left(\tilde{\sigma}_{F}^{2}+\tilde{\sigma}_{B}^{2}\right)^{3/2}}+\frac{\sqrt{2}S\left(\tilde{a}_{11}+\tilde{a}_{22}+\tilde{a}_{12}\right)}{\tilde{\sigma}_{B}^{2}}\right]=0
\end{array}\right.\label{eq:sdeviations}
\end{split}
\end{equation}
As discussed in the main text and above one takes $a_{12}=2a_{\mathrm{scat}}$
and $a_{11}=a_{22}=a_{\mathrm{scat}}$. Due to the difference on the interaction terms of the two equations of (\ref{eq:sdeviations}) 
the assumption $a_{1F}=a_{2F}=2a_{\mathrm{scat}}$
does not automatically satisfy 
the requirement of the Hamiltonian parameters of the
proposal. For $S$ small, at least, the result is approximately valid
so these values are also taken as reference for the scattering between
bosons and fermions. Equations (\ref{eq:sdeviations}) are used to obtain the 
data reported in Figure \ref{fig:plots}.

An important check concerns whether the values of the
parameters validate the perturbative approximation obtained for large values 
of $U$. This amounts to
check that $t_{\alpha}/U$ and $V_{0\alpha}\Delta_{\alpha}$ are actually 
small (here it is considered that they should be $\sim0.1$ or smaller).
For illustrative purposes it was fixed $S=1$, $V_{0F}=V_{0B}=V_{0}$ and
$\Delta_{F}=\Delta_{B}=\Delta$. 
Direct analysis of the above equations shows that
in order to guarantee that the perturbative regime is valid in 
the interval $\tilde{V}_{0}\sim3-10$, then one should have 
$\tilde{\sigma}_{\perp}\sim0.2$
and $\Delta\lesssim10^{-3}$. If one takes $\tilde{\sigma}_{\perp}$
to be, say, two or three times higher than this, larger potential amplitudes
are required. Alternatively, larger scattering lengths could also be
used. From the other side, there is some freedom in choosing
the values of $\Delta$ in order to remain in the perturbative regime.
However this choice should respect the fact the two minima 
should still be present at $x=\pm a/2$ which is translated 
into $\left|\Delta\right|<1/3$.
Finally, the mass parameter of the target model will
scale as $V_{0}\Delta$ and the choice was taken such that the energy
scale of this term matches the order of magnitude of the other terms
in the Hamiltonian $t_{\alpha}t_{\beta}/U\sim V_{0}\Delta$, which
results to be $\Delta\lesssim10^{-3}$, as referred in the main text.

\section{Equations of motion and energy-momentum tensor for theories with
higher derivatives}\label{sec:EMtensor}

Here a classical field theory with higher derivatives is considered.
The well known Euler-Lagrange equations are derived by the extremization
of the action. The inclusion of higher derivatives on the Lagrangian
lead to a reformulation of the equations. In fact by calculating explicitly
$\delta S=0$, integrating by parts whenever necessary one obtains:
\begin{equation}
\underset{n=0}{\overset{N}{\sum}}\left(-1\right)^{n}\partial_{\mu_{1}}\ldots\partial_{\mu_{n}}\frac{\partial{\cal L}}{\partial\left(\partial_{\mu_{1}}\ldots\partial_{\mu_{n}}\phi\right)}=0,\label{eom}
\end{equation}
where $N$ is the highest number of derivatives appearing in a term
of the Lagrangian. For $N=1$ one recovers the usual Euler-Lagrange
equations. 

Consider now a general translation $x^{\mu}\rightarrow x^{\mu}+\varepsilon^{\mu}$.
The total change of the Lagrangian is 
\begin{equation}
\delta{\cal L}=\frac{\delta{\cal L}}{\delta\left(\partial_{\mu_{1}}\ldots\partial_{\mu_{n}}\phi\right)}\delta\left(\partial_{\mu_{1}}\ldots\partial_{\mu_{n}}\phi\right).
\end{equation}
which results in 
\begin{equation}
\delta{\cal L}=\frac{\partial{\cal L}}{\partial\left(\partial_{\mu_{1}}\ldots\partial_{\mu_{n}}\phi\right)}\partial_{\nu}\partial_{\mu_{1}}\ldots\partial_{\mu_{n}}\phi\varepsilon^{\nu}.
\end{equation}
The derivatives can be written as acting on $\frac{\partial{\cal L}}{\partial\left(\partial_{\mu_{1}}\ldots\partial_{\mu_{n}}\phi\right)}$
with a minus sign plus a total derivative term:
$$
\frac{\partial{\cal L}}{\partial\left(\partial_{\mu_{1}}\ldots\partial_{\mu_{n}}\phi\right)}\partial_{\nu}\partial_{\mu_{1}}\ldots\partial_{\mu_{n}}\phi
$$
$$
=\partial_{\mu_{1}}\left(\frac{\partial{\cal L}}{\partial\left(\partial_{\mu_{1}}\ldots\partial_{\mu_{n}}\phi\right)}\partial_{\mu_{2}}\ldots\partial_{\mu_{n}}\partial_{\nu}\phi\right)
$$
\begin{equation}
-\partial_{\mu_{1}}\frac{\partial{\cal L}}{\partial\left(\partial_{\mu_{1}}\ldots\partial_{\mu_{n}}\phi\right)}\partial_{\mu_{2}}\ldots\partial_{\mu_{n}}\partial_{\nu}\phi
\end{equation}
By continuing this process with every $\partial_{\mu_{i}}$ acting
on $\phi$ in the terms that are not an exact derivative (last term),
one obtains:
$$
\frac{\partial{\cal L}}{\partial\left(\vec{\partial}\phi\right)}\partial_{\nu}\partial_{\mu_{1}}\ldots\partial_{\mu_{n}}\phi
$$
$$
=\underset{i=1}{\overset{n}{\sum}}\left(-1\right)^{i-1}\partial_{\mu_{i}}\left(\partial_{\mu_{1}}\ldots\partial_{\mu_{i-1}}\frac{\partial{\cal L}}
{\partial\left(\vec{\partial}\phi\right)}
\partial_{\mu_{i+1}}\ldots\partial_{\mu_{n}}\partial_{\nu}\phi\right)$$
\begin{equation}
+\left(-1\right)^{n}\partial_{\mu_{1}}\ldots\partial_{\mu_{n}}
\frac{\partial{\cal L}}
{\partial\left(\vec{\partial}\phi\right)}
\partial_{\nu}\phi,
\end{equation}
where, in order to simplify the notation, it was written
$$\frac{\partial{\cal L}}{\partial\left(\vec{\partial}\phi\right)} \equiv \frac{\partial{\cal L}}{\partial\left(\partial_{\mu_{1}}\ldots\partial_{\mu_{n}}\phi\right)}.$$

The special case of $n=0$ 
gives $\frac{\partial{\cal L}}{\partial\phi}\partial_{\nu}\phi$.
Summing over all $n$, joining all terms coming from the part of
the above expression one can recognize the equations of motion (\ref{eom})
and therefore they are put to zero. What remains is a series of total
derivatives. Furthermore in itself the Lagrange density changes as 
$\delta{\cal L}=\varepsilon^{\nu}\partial_{\nu}\phi$. By rearranging
the variables one obtains:
$$
\underset{n=1}{\overset{N}{\sum}}\underset{i=1}{\overset{n}{\sum}}\left(-1\right)^{i-1}\partial_{\mu_{1}}\left(\partial_{\mu_{2}}\ldots\partial_{\mu_{i}}\frac{\partial{\cal L}}
{\partial\left(\vec{\partial}\phi\right)}
\partial_{\mu_{i+1}}\ldots\partial_{\mu_{n}}\partial_{\nu}\phi\right)\varepsilon^{\nu}
$$
\begin{equation}
-\partial_{\mu_{1}}{\cal L}\delta_{\nu}^{\mu_{1}}\varepsilon^{\nu}=0
\end{equation}
and therefore one can identify the energy-momentum tensor. This is just
the conserved current that follows from Noether's theorem for the
special case of space-time translations:
$$T^{\mu\nu}=\underset{n=0}{\overset{N}{\sum}}\underset{i=0}{\overset{n}{\sum}}\left(-1\right)^{i}\partial_{\mu_{1}}\ldots\partial_{\mu_{i}}\frac{\partial{\cal L}}{\partial\left(\partial_{\mu}\partial_{\mu_{1}}\ldots\partial_{\mu_{n}}\phi\right)}\partial_{\mu_{i+1}}\ldots$$
\begin{equation}
\ldots\partial_{\mu_{n}}\partial^{\nu}\phi-
{\cal L}\eta^{\mu\nu}.\label{eq:en-mom_tensor}
\end{equation}

\clearpage

\end{document}